\documentclass{emulateapj}  % for emulate style
\usepackage{epsf}
\usepackage{latexsym}  % for emulate style
\usepackage{dcolumn}% Align table columns on decimal point
\input{epsf}
\usepackage{epsfig}
\usepackage{natbib}

\begin{document}

\bibstyle{aas}

\def\msun{{\rm M}_\odot}

\shorttitle{Scatter in the Y-M relation}
\shortauthors{Shaw et al.}

\author{Laurie D. Shaw and Gilbert P. Holder}
\affil{Department of Physics, McGill University, Montreal QC H3A 2T8}
\and
\author{Paul Bode }
\affil{Princeton University Observatory, Princeton, NJ 08544-1001}
\email{lds@physics.mcgill.ca}

\title{The Impact of halo properties, energy feedback and projection
effects on the Mass-SZ Flux relation}

\begin{abstract}
We present a detailed analysis of the intrinsic scatter in the
integrated SZ effect - cluster mass ($Y-M$) relation, using
semi-analytic and simulated cluster samples.  Specifically, we
investigate the impact on the $Y-M$ relation of energy feedback,
variations in the host halo concentration and substructure
populations, and projection effects due to unresolved clusters along
the line of sight (the SZ background).  Furthermore, we investigate at
what radius (or overdensity) one should {\it measure} the integrated
SZE and {\it define} cluster mass so as to achieve the tightest
possible scaling. We find that the measure of $Y$ with the least scatter
is always obtained within a smaller radius than that at which the mass
is defined; e.g. for $M_{200}$ ($M_{500}$) the scatter is least for
$Y_{500}$ ($Y_{1100}$). The inclusion of energy feedback in the gas
model significantly increases the intrinsic scatter in the $Y-M$
relation due to larger variations in the gas mass fraction compared to
models without feedback. We also find that variations in halo
concentration for clusters of a given mass may partly explain why the
integrated SZE provides a better mass proxy than the central
decrement. Substructure is found to account for approximately 20\% of
the observed scatter in the $Y-M$ relation. Above $M_{200} = 2\times
10^{14} h^{-1} \msun$, the SZ background does not significantly effect
cluster mass measurements; below this mass, variations in the
background signal reduce the optimal angular radius within which one
should measure $Y$ to achieve the tightest scaling with $M_{200}$.
\end{abstract}

\section{INTRODUCTION}

The evolution of the number density of galaxy clusters is a sensitive
cosmological probe \citep{bahcall:98,Eke:98b}.  As an indicator of the
expansion rate as a function of time, the galaxy cluster number
density is sensitive to the dark energy equation of state
\citep{haiman:00,weller:01}.  This provides a growth-based dark energy
test, an important complement to the distance-based tests that have
provided the most compelling evidence for dark energy to this point
\citep{perlmutter:99,schmidt:98}.  Galaxy clusters can be selected by
many diverse methods, including (but not limited to) optical richness,
X-ray thermal bremsstrahlung flux, and weak lensing shear.  The
Sunyaev-Zeldovich (SZ) effect \citep{sunyaev:72, birkinshaw:99,
carlstrom:02} is being actively pursued as a method to detect galaxy
clusters to high redshift, and experiments such as the Atacama
Cosmology Telescope \citep{Kosowsky:03} and the South Pole Telescope
\citep{Ruhl:04} are currently surveying the microwave sky to develop
large catalogs of galaxy clusters that are uniformly selected.

The key challenge for using galaxy clusters as precise cosmological
probes is in understanding how to relate observables (SZ flux, X-ray
flux, optical richness, weak lensing shear, etc.) to a quantity that
can be well predicted by theory, namely mass. The ultimate goal is to
produce theoretical predictions of the distributions of observables as
a function of redshift and cosmological parameters. Short of this, one
approach is to theoretically model the evolution of number density as
a function of mass, and then estimate the mapping between observables
and mass in order to predict the observed evolution.  This mapping can
either be estimated from theoretical considerations or be determined
directly from the data, assuming some regularity in the mapping
\citep{majumdar:03, hu:03, lima:04, lima:05}.  In either case, it is
also important to understand the {\em scatter} in the mapping between
observable and mass \citep{lima:05}.

One reason that SZ surveys are an attractive possibility for dark
energy studies is that the scatter in the mapping between SZ flux and
mass is expected to be small from straightforward physical
considerations \citep{barbosa:96}: the integrated SZ flux is a direct
indicator of the total thermal energy in the intracluster medium. It
is exceedingly difficult to disrupt the global energy budget on this
scale, as the total thermal energy of a galaxy cluster is on the order
of $10^{62-63}$ ergs.  This expectation has been verified by
hydrodynamical simulations of the SZ effect (SZE) \citep{Nagai:06, Motl:05,
White:02}.

It is important to identify sources of scatter; even though the
scatter is expected to be small, it is still non-negligible
\citep{lima:05}.  Furthermore, the SZ flux integrated to infinity is
not an observable, and the definition of the mass of a halo is
ambiguous in a CDM paradigm \citep{white:01}.  One must carefully
define SZ flux and mass within some fiducial region.  The mapping
between SZ flux and mass could plausibly be more or less tight
(i.e. have more or less scatter) for different definitions of SZ flux
and mass. In this paper we will address this question.

The key physical processes that govern the density and temperature
profiles of the hot baryonic gas in cluster interiors are still poorly
understood and improving this situation is currently a very active
area of research \citep[see ][ and references therein]{Motl:05,
Borgani:05, Nagai:06, Ettori:06, Sijacki:06, Romeo:06,
Muanwong:06}. Dark matter halos provide the potential well into which
cluster baryons sink, setting the gas temperature scale.  However, in
order for theoretical work to be reconciled with the observations, it
is clear that non-gravitational processes such as radiative cooling,
star-formation and energy feedback (through supernovae, AGN outflows
and galactic winds) play an important role \citep{Evrard:91,
Kaiser:91, Balogh:99, Eke:98, McCarthy:03, Solanes:05,
Bode:07}. Unfortunately, many of these processes involve small-scale,
sub-grid physics that are very difficult to simulate
directly. Typically a semi-analytic prescription must be incorporated
in simulations to approximate these effects.

Variations in the internal dynamics between clusters will be
a significant contributor to the intrinsic scatter in the mass--SZ
flux scaling relation.  Furthermore, the large scatter in the halo
mass-concentration relation \citep{Jing:00, Eke:01, Bullock:01b,
Dolag:04, Shaw:06, Maccio:07} and the wide range in substructure
populations \citep{DeLucia:04, Gao:04, Gill:04b, Shaw:07} results in a
gravitational framework that can vary significantly between clusters
of similar mass and redshift. In this study, we apply several
different realisations of a semi-analytic model of intracluster gas,
calibrated using observations, to the output of a high-resolution
N-body lightcone simulation to investigate some of these questions.

Another significant contributor to the scatter in the SZ flux-mass
scaling relation is errors in flux measurement caused by confusion due
to projection effects \citep{Holder:07}. \citet{Motl:05} demonstrated
that the thermal SZE integrated over a large fraction of the projected
virial region of the cluster allows a robust measurement of cluster
mass, suppressing the impact of heating and cooling in the
core. However, the noise due to background clusters will increase as
one increases the angular size of the aperture within which the SZE is
measured.  It is important to assess the impact of these interlopers,
particularly those clusters that are too dim to be identified
individually and accounted for. In this study we investigate the
angular scales at which scatter in the $Y-M$ relation due to projection
errors begins to dominate over that due to variations in internal
cluster properties.

The outline of this paper is as follows. In Section \ref{sec:method}
we provide a brief overview of the mass-flux relation, and describe
the simulations and gas model that we use to predict this relation. In
Section \ref{sec:results} we investigate the intrinsic scatter in
this relation due to the choice of cluster mass definition, the area
within which the SZE flux is measured, the impact of variations in
cluster gas temperature and density, and the variations in host halo
concentration and substructure populations. In Section
\ref{sec:projection} we use a simulated
lightcone to investigate the impact of projection effects on mass
estimates as a function of cluster and aperture angular size. Finally,
in Section \ref{sec:discussion} we summarise and discuss our
conclusions.

\section{Method}\label{sec:method}

\subsection{Sunyaev-Zel'dovich Scaling Relations}

The thermal SZE is a distortion of the CMB caused by inverse Compton
scattering of CMB photons (at temperature $T_{cmb}$)
off electrons (at $T_e$) in the high temperature plasma
within galaxy clusters. To first order, the temperature change 
at frequency $\nu$ of the CMB 
is given by 
$\Delta T_{\nu}/T_{cmb} = f_{\nu}(x) y$, 
where $f_{\nu}(x) = x(\coth(x/2) - 4)$, $x = h\nu / k_B T_{cmb}$, 
and $y$ is the normal Compton parameter
\begin{equation}
\frac{\Delta T_{cmb}}{T_{cmb}} \equiv y = \left(\frac{k_B \sigma_T}{m_e
c^2}\right)\int n_e(l)T_e(l) dl \;,
\end{equation}
where the integral is along the line of sight. The {\it integrated}
temperature distortion (the SZ flux) across the surface of a cluster
is defined as
\begin{equation}
Y_z = \int y d\Omega \;, 
\end{equation}
where $\Omega$ is the solid angle subtended by the cluster on the
sky. For a cluster at redshift $z$,
\begin{equation}
Y_z = \frac{1}{d^2_A(z)}\left(\frac{k_B \sigma_T}{m_e
c^2}\right)\int n_e(l)T_e(l) dV \;,
\label{eqn:integratedy}
\end{equation}
where $d_A(z)$ is the angular diameter distance, and the integral is
now over the volume of the cluster. $Y_z$ is thus measured in angular
units squared. However, it is often assumed that simulated clusters
reside at a redshift of zero, and thus the $1/d_A(z)^2$ factor is
omitted. In this case the units of $Y$ are Mpc$^2$ (henceforth
represented by omitting the subscript $z$). In the following section,
in which we investigate the intrinsic scatter in the $Y-M$ relation
due to variations in internal cluster properties, we adopt the latter
definition. In the final section, where we investigate the impact of
projection effects on the $Y_z-M$ relation using a simulated
lightcone, we measure the integrated Compton parameter as defined in
Eqn. \ref{eqn:integratedy}.

The mass of a halo is typically defined as the mass contained within a
region of spherical overdensity $\Delta$ times greater than the
critical density,
\begin{equation}
M_{\Delta} = \frac{4}{3}\pi R_{\Delta}^3 \Delta \rho_c(z) \;,
\label{eqn:mdelta}
\end{equation}
where $\rho_c(z)$ is the critical density at redshift $z$.

From spherical collapse theory, the region in which matter is
predicted to have virialized is defined at $z=0$ by the overdensity
contour $\Delta_c=178$ in a $\Omega_M=1$ universe or $\Delta_c=96$ in
a flat $\Omega_M=0.26$ universe \citep{Lahav:91, Lacey:93, Cole:96,
Eke:96, Bryan:98}.  Most recent studies of halos in simulations define
mass adopting either the redshift dependent $\Delta_c$, or a fixed
overdensity factor of $\Delta = 200$. \citet{Evrard:07} compare common
definitions of halo mass using a suite of N-body simulations, focusing
specifically on the the $M-\sigma_{\rm DM}$ virial relation, where
$\sigma_{\rm DM}$ is the one-dimension dark matter velocity
dispersion. They constrain the region of minimum variance in the
virial relation to being within the overdensity contour $\Delta =
200$.

In the absence of non-thermal heating and cooling processes, we expect
a self-similar scaling between cluster mass and SZ flux:
\begin{equation}
Y \propto f_{\rm gas}M^{(5/3)} E(z)^{2/3} \;,
\label{eqn:selfsimilar}
\end{equation}
where $f_{\rm gas}$ is the cluster gas fraction, and $E(z) = (\Omega_{\rm
M}(1+z)^3 + \Omega_\Lambda)^{1/2}$. In practice, many studies of
simulated clusters have measured a steeper slope than $(5/3)$,
attributed to the presence of radiative cooling and non-gravitational
heating in their simulations \citep{White:02, daSilva:04, Motl:05,
Nagai:06}.

To date, there have been only a limited number of clusters observed
using the SZE. Recently, \citet{Morandi:07}, in an X-ray and SZ study
of 24 clusters in the redshift range 0.14-0.82, found a good agreement
between the slope of the observed $Y-M$ relation and the self-similar
predictions for their cooling core sample.  However, they find a
significantly shallower slope when they include both cooling-core and
non cooling-core clusters in their analysis \citep[see also
][]{Benson:04}. They also measure greater scatter in the $Y-M$ relation
than is typically seen in simulations.

The existence of a tight SZ flux-mass scaling relation provides the
crucial link between SZ cluster surveys and constraining cosmological
parameters. It is therefore important to understand the character and
physical origins of the scatter in this relationship,
and how one can suppress its impact to
obtain an accurate measure of cluster masses. In
particular, there has been little investigation concerning at what
radius (or equivalently, what overdensity) one should attempt to {\it
measure} the integrated SZ flux, and {\it define} cluster mass, so as
to achieve the tightest possible scaling. We investigate this issue in
detail in Section \ref{sec:results}. 

\subsection{Simulations and Gas Model}
\label{sec:gasmod}
For this study we adopt two complementary approaches to measuring and
characterising the scatter in the $Y_{\Delta}-M_{\Delta}$ scaling
relation. Our main cluster samples are generated through applying
different realisations of an intracluster gas model 
\citep{Ostriker:05, Bode:07} to individual dark
matter halos identified in a high resolution N-body simulation.
For comparison, we also use a second sample of clusters
taken from the output of an SPH simulation. In this section we
describe in more detail the simulations and gas model used in this
analysis, and the derived cluster catalogs.

To create our main simulated cluster catalog, we begin with the
output of a large ($N=1024^3$ particles) cosmological dark matter
simulation.  The cosmology was chosen to be consistent with
that measured from the 3rd-year WMAP data combined with large-scale
structure observations \citep{Spergel:07}, namely a spatially
flat LCDM model with parameters: baryon density $\Omega_b=0.044$; 
total matter density $\Omega_m=0.26$; cosmological constant
$\Omega_\Lambda=0.74$; linear matter power spectrum amplitude
$\sigma_{8}=0.77$; primordial scalar spectral index $n_s=0.95$; and
Hubble constant ${\rm H}_{0}=72 {\rm km\,s^{-1}Mpc^{-1}}$
(i.e. $h=0.72={\rm H}_{0}/100{\rm km\,s^{-1}Mpc^{-1}}$).  The
simulated volume is a periodic cube of size $L=320h^{-1}$Mpc; the
particle mass $m_p=2.2\times 10^{9}h^{-1}M_\odot$, and the cubic spline
softening length $\epsilon=3.2h^{-1}$kpc.

The matter distribution in a light cone covering one octant of the sky
extending to $z=0.5$ was saved in 315 time slices. Dark matter halos
were identified using the Friends-of-Friends algorithm with a comoving
linking length parameter $b=0.2$. For analysis presented here, we
select a sample containing the 1267 lowest redshift clusters of mass
$M_{vir} \geq 5\times 10^{13} h^{-1} \msun$, spanning a redshift range
of $0.002 \leq z \leq 0.14$.

The cluster gas distribution in each halo was calculated using the
semi-analytic model described in
\cite{Ostriker:05} and \citet{Bode:07}.  In
brief, a 3D mesh (with cell side-length $4\epsilon = 12.8 h^{-1}$ kpc)
is placed around each cluster, with the gas pressure and density
determined in each mesh cell assuming hydrostatic equilibrium and a
polytropic equation of state \citep[with adiabatic index
$\Gamma=1.2$,][]{Ascasibar:06}.  It is also assumed that the gas in the
densest cluster regions has cooled and condensed, forming stars. At z
= 0, the stellar/gas mass ratio is set to 0.1 \citep{Lin:03,
Voevodkin:04}. To compute the star/gas ratio at $z > 0$, the star
formation rate was assumed to follow a delayed exponential model
\citep[Eqn. 1 of][]{Nagamine:06}, with decay time $\tau=1.5$Gyr.

As discussed in detail in \cite{Bode:07}, the most important free
parameter in this model is the energy input into the cluster gas via
nonthermal feedback processes, such as AGN outflows and supernovae.
This is set through the parameter $\epsilon_{f}$, such that the
feedback energy is $\epsilon_{f}M_*c^2$, where $M_*$ is the stellar
mass in the cluster. In this study, we set $\epsilon_{f} = 4\times
10^{-5}$, which is a value that provides good agreement between
simulation cluster properties and observed X-ray scaling relations
\citep[see Fig. 2 of][]{Bode:07}.

To investigate the impact of the feedback parameter and the assumption
of hydrostatic equilibrium on the $Y-M$ scaling relation, we have
created three realisations of our cluster sample. In the first, the
gas distribution in each halo is calculated using the full model (for
which the measured X-ray scaling relations agree well with
observations). For the second sample we remove the effects of feedback
by setting $\epsilon_f$ to zero, so that the gas is merely re-arranged
into hydrostatic equilibrium having removed the fraction that is
assumed to have formed stars.

For our final sample, we take an very simplistic approach, assuming an
isothermal model in which the gas directly follows the dark matter
distribution. This is achieved through setting the gas density in each
mesh cell to $\Omega_b/\Omega_{DM}$ times the interpolated dark matter
density. The gas temperature is constant with radius, with the global
temperature set according to the scaling relation measured by
\citet{Vikhlinin:06},
\begin{equation}
E(z) M_{500} h^{-1} \msun = A_X \left( \frac{kT_{mw}}{5 {\rm keV}} \right)^\alpha \;,
\label{eqn:scalinglaw}
\end{equation}
where $A_X = 3.32 \times 10^{14} h^{-1} \msun$, $T_{mw}$ is the gas mass
weighted temperature, and the slope $\alpha = 1.47$. We emphasize that
this final sample does not represent a physically accurate model, and
is included as an aid to understand the results of the more
sophisticated approach.  Hereafter, we refer to each sample as FULL,
NOFB and BASIC, respectively.

In practice, $Y$ is measured for the clusters in each of our samples
by summing up the electron pressure $n_e T_e$ in each mesh cell along
one direction, thus creating a 2-d $y$ image of the cluster \citep[for
example, see Fig. 7 of][]{Ostriker:05}.  Note that this is directly
equivalent to an integral along the line-of-sight through the
cluster. We then calculate the integrated $Y$ parameter, by summing all
cells that lie within circular aperture of radius $R_\Delta$ (defined
below) from the cluster centre.

\subsection{Hydrosimulation}

For comparison with the gas model results, we also make use of a
(publicly available\footnote[1]{taken from
\url{http://virgo.susx.ac.uk/clusdata.html}}) sample of clusters from
the output of an smoothed-particle-hydrodynamics (SPH) simulation,
evolved using the HYDRA code \cite{Couchman:95, Pearce:97}. The
simulation contained $160^3$ gas and $160^3$ dark matter particles in
a box of side length $100 h^{-1}$Mpc. The cosmological parameters were
$\Omega_{M} = 0.35$, $\Omega_b = 0.038$, $\Omega_\Lambda = 0.65$,
$\sigma_8 = 0.9$ and $h = 0.71$. The gravitational softening length
was set to $25 h^{-1}$ kpc, once the simulation had evolved past a
redshift of 1. The simulation is described in more detail in
\citet{Muanwong:01}.

It is important to note that this simulation did not contain any
radiative cooling and so the dense, cold gas, in the centre of
clusters is unable to condense and form stars. 
\citet{Muanwong:01, Muanwong:02} and \citet{Thomas:02} 
find that the inclusion of radiative cooling
removes this gas and heats the surrounding material, providing a
temperature and entropy boost to the ICM. Inclusion of radiative
cooling was also found to be necessary in order to match observed
X-ray scaling relations. Therefore, the cluster gas in the SPH
simulation used here is likely to be colder and significantly more
centrally concentrated than expected in more physically realistic
models.

Cluster particles are selected by identifying all particles within a
cube of side-length $4 R_{vir}$ around each cluster centre in the $z =
0$ simulation snapshot, where the virial radius and cluster centre are
taken from the cluster catalog of \citet{Muanwong:01}. In total, our
sample contains 212 clusters with a minimum virial mass of $2.5\times
10^{13} h^{-1} \msun$ (where the virial overdensity $\Delta_v \approx
108$).  All clusters contain at least 1000 particles (dark matter plus
gas) within their virial radius.

To measure the integrated SZ flux we place a 3d mesh around each
cluster, calculating the interpolated gas density and temperature
within each cell according to the smoothing kernel given in
\citet{Thomas:92}. For consistency we set the mesh cell side-length equal
to that used in the gas model described in the previous section ($12.8
h^{-1}$ kpc). The integrated SZ $Y$ parameter is then calculated for each
cluster from the mesh in exactly the same way as for the gas model
samples.  Henceforth, we label this sample SPH.

\section{Results}
\label{sec:results}
\subsection{Measuring $Y-M$ scaling relations}

For each cluster sample, we measure both the mass $M_{\Delta}$ and the
integrated Compton $Y$ parameter $Y_{\Delta}$ within regions of density
$\Delta \rho_c$, where the overdensity varies within the range $50
\leq \Delta \leq 1500$ -- corresponding to roughly $0.37 \leq
r/R_{200} \leq 1.77$ -- in steps of $\Delta = 100$. Thus for each
cluster we obtain 16 measures each of $M$ and $Y$. From these various
measures we will determine empirically the combination of $M_{\Delta}$
and $Y_{\Delta}$ that provides the lowest scatter around a power-law
relation. The outermost value of $\Delta = 50$ is outside the virial
overdensity calculated from spherical collapse, $\Delta_c \approx
95$. The innermost value of $\Delta = 1500$ was chosen for two
reasons. First, we are constrained by the size of the mesh cells in
the gas model. Second, none of our cluster samples are expected to
correctly reproduce the gas temperature and density in the cluster
core ($r \leq 0.1R_{200}$), due to the complicated interplay between
cooling and heating processes in these regions. Hence we pick an inner
boundary which is considerably outside the cluster core.

For each of the 256 combinations of $\Delta_{\rm Y}$ and 
$\Delta_{\rm M}$, we fit their relation with a power-law,
\begin{equation}
\hat{Y} = E(z)^{2/3} 10^{A} \left(\frac{M_{\Delta}}{10^{14} h^{-1}\msun}\right)^{\alpha} \; ,
\label{eqn:power-law}
\end{equation}
where $10^{A}$ is the normalisation at $10^{14} h^{-1} \msun$ and
$\alpha$ is the slope. We define the scatter around the best fit
relation as;
\begin{equation}
\sigma_{\rm YM} = \left( \frac{\sum_{i=1}^{N}(\ln Y_{\Delta_{\rm Y}}
- \ln \hat{Y}_{\Delta_{\rm M}})_i^2}{N-2} \right)^{1/2} \;,
\label{eqn:scat}
\end{equation}
where $Y_{\Delta_{\rm Y}}$ is the flux measured within $R_{\Delta_{\rm
Y}}$, $\hat{Y}_{\Delta_{\rm M}}$ is the fitted flux for cluster of
mass $M_{\Delta_{M}}$ and N is the total number of clusters in the
sample.  The fitting is performed using a
Levenberg-Marquardt algorithm. To avoid introducing a bias into our
results by having defined the minimum cluster mass in our sample at a
particular overdensity, we fit (and calculate $\sigma_{\rm YM}$) using
only the 1000 most massive clusters at each $\Delta_{\rm M}$ in our
gas code samples. This corresponds to a limiting mass of $5\times 
10^{13} h^{-1} \msun$ at $\Delta_{\rm M} = 200$. For the SPH sample,
we take the 150 most massive clusters, corresponding to a minimum mass
of $3 \times 10^{13} h^{-1} \msun$ at $\Delta_{\rm M} = 200$.

We note that for all combinations of $\Delta_{\rm Y}$ and $\Delta_{\rm
M}$, a power-law is found to be a very good fit to the $Y-M$
relation. The scatter, $\delta = \ln Y - \ln \hat{Y}$, around this
relation in the vertical direction is gaussian with a slight tail
towards low (high) values of $Y$ for low (high) $\Delta_M$.  Hence the
scatter, $\sigma_{\rm YM}$, in Eqn. \ref{eqn:scat} represents the
standard deviation of the distribution of $\ln Y_{\Delta_{\rm
Y}}$ around the power-law fit.

\subsection{Slope and Normalisation}

\begin{table*}
\begin{center}
\begin{tabular}{c c c c c c c c c}
\hline
\hline
 & & & & & & & & \\
$\Delta_M$ (=$\Delta_Y)$ & $\alpha_{FB}$ & $A_{FB}$  & $\alpha_{NOFB}$ & $A_{NOFB}$ & $\alpha_{Basic}$ & $A_{Basic}$ & $\alpha_{SPH}$ & $A_{SPH}$\\
 & &  & & & & & & \\
\hline
 & &  & & & & & & \\
50 & 1.51 ($\pm 0.05$)  &-5.59 ($\pm 0.01$) & 1.49 ($\pm 0.03$) & -5.52 ($\pm 0.01$) & 1.64 ($\pm 0.02$) & -5.59 ($\pm 0.01$) & 1.64 ($\pm 0.05$)  & -5.71  ($\pm0.01$) \\
 & &  & & & & & & \\
100 & 1.62 ($\pm 0.05$)& -5.64 ($\pm 0.01$) & 1.51 ($\pm 0.03$) & -5.50  ($\pm 0.01$) & 1.64  ($\pm 0.02$) & -5.54 ($\pm 0.01$) & 1.64 ($\pm 0.05$)  & -5.63  ($\pm0.01$) \\
&  &  & & & & & & \\
200  & 1.70 ($\pm 0.05$)&-5.62 ($\pm 0.01$) & 1.51 ($\pm 0.04$) & -5.44 ($\pm 0.01$) & 1.65 ($\pm 0.02$) & -5.46 ($\pm 0.01$) & 1.67 ($\pm 0.06$)  & -5.54  ($\pm0.02$) \\
& &  & & & & & & \\
500  & 1.83 ($\pm 0.05$)&-5.50 ($\pm 0.01$) & 1.53 ($\pm 0.03$) & -5.30 ($\pm 0.01$) & 1.66  ($\pm 0.02$) & -5.29 ($\pm 0.01$) & 1.72 ($\pm 0.05$)  & -5.39  ($\pm0.02$) \\
& &  & & & & & & \\
1000 & 1.92 ($\pm 0.04$)&-5.36 ($\pm 0.01$) & 1.55 ($\pm 0.03$)& -5.19 ($\pm 0.01$) & 1.65 ($\pm 0.02$) & -5.16 ($\pm 0.01$) & 1.77 ($\pm 0.04$)  & -5.25  ($\pm0.02$) \\
& &  & & & & & & \\
1500 & 1.98 ($\pm 0.04$)&-5.26 ($\pm 0.02$) & 1.57 ($\pm 0.03$) & -5.12 ($\pm 0.01$) & 1.56 ($\pm 0.03$) & -5.10 ($\pm 0.01$) & 1.77 ($\pm 0.04$)  & -5.17  ($\pm0.03$) \\
 & &  & & & & & & \\
\hline
\end{tabular}
\caption{Slope $\alpha$ and normalisation index $A$ measured for
a selection of overdensities for each cluster sample.}
\label{tab:slopenorm}
\end{center} 
\end{table*}

To enable comparison of the results of our gas model and SPH
simulation clusters with previous work, in Table \ref{tab:slopenorm}
we give the slope $\alpha$ and normalisation index $A$ (and their
associated errors) from the power-law fit to the $Y-M$ relation at
selected values of $\Delta_{\rm Y} = \Delta_{\rm M}$.

At $\Delta = 200$, the slope for the FULL cluster sample (1.70) is
close to the $5/3$ expected from spherical collapse theory. Within
smaller radii, the slope becomes steeper. This is due to the
increasing impact of energy feedback on the gas fraction, $f_{\rm gas} =
M_{\rm g}(<R)/M_{\rm tot}(<R)$ within decreasing $R$.  The
additional energy increases the gas temperature, inflating the overall
gas distribution and decreasing the baryon fraction within any given
radius.  For lower mass clusters this effect is more pronounced, thus
steepening the slope of the $Y-M$ relation 
\citep[see Figure 2 of][]{Bode:07}.
The slope $\alpha$ becomes greater than 5/3 due to
the weak mass dependence of the gas mass fraction. This mass
dependence has been observed in hydrodynamical simulations
(\citet{Muanwong:02, daSilva:04, Kravtsov:05}, though see
\citet{Ettori:06}). We refer the reader to \citet{Bode:07} for a more
extensive discussion on the gas fraction obtained from this model,
and comparisons to other studies and observations.

In general, both the slope and normalisation that we measure for this
model are in reasonably good agreement with the results of
\citet{Nagai:06} for their hydrodynamic simulations including cooling,
star formation, and supernova feedback \citep{Kravtsov:02,
Kravtsov:05}, differing by less than 6\% in $\alpha$ and less than 1\%
in $A$ at $\Delta = 200$ and $500$. \citet{daSilva:04} investigated
the impact of including radiative cooling or pre-heating the gas in
their simulations on the $Y_{200}-M_{200}$ relation, finding slopes of
1.69, 1.79 and 1.93 for their adiabatic, cooling and preheating runs
respectively (their adiabatic simulation is the same as that used in
this study). They also find that the increased slope is due to a
decrease in the gas fraction in lower mass ($M_{200} < 10^{14} h^{-1}
\msun$) clusters in the the cooling and preheating simulations. The
steeper slopes measured compared to this study are probably due in part
to the lower mass threshold $10^{13} h^{-1} \msun$ in their cluster
samples.  \citet{Motl:05} measure an increase of 0.1 in slope between
their adiabatic and cooling simulations, although the slope decreased
again once they allowed for star-formation in their simulations.

The slopes measured from the NOFB sample are typically much lower than
those of our other samples or in previous studies, indicating that
assumptions of hydrostatic equilibrium alone (i.e. without feedback)
do not reproduce the results of more sophisticated
simulations. Indeed, more realistic slopes are measured from our BASIC
sample, although this is likely due to the way in which cluster
temperature in this sample is calibrated from observational results.

The results for our SPH sample match closely those of previous
adiabatic simulations, although we note the lower normalisation
measured here compared to the adiabatic simulations of
\citet{Nagai:06} ($A = -5.54$ compared to -5.42 at $R_{200}$,
corresponding to a decrease of 25\% in Y); the difference is probably
due to the different ratios of $\Omega_b/\Omega_M$ between the
simulations (0.11 and 0.14, respectively). This demonstrates the
sensitive dependence of the normalisation of the $Y-M$ relation on
$\Omega_b/\Omega_M$.

\subsection{Scatter in $Y-M$ relation}

\begin{figure*}
\plotone{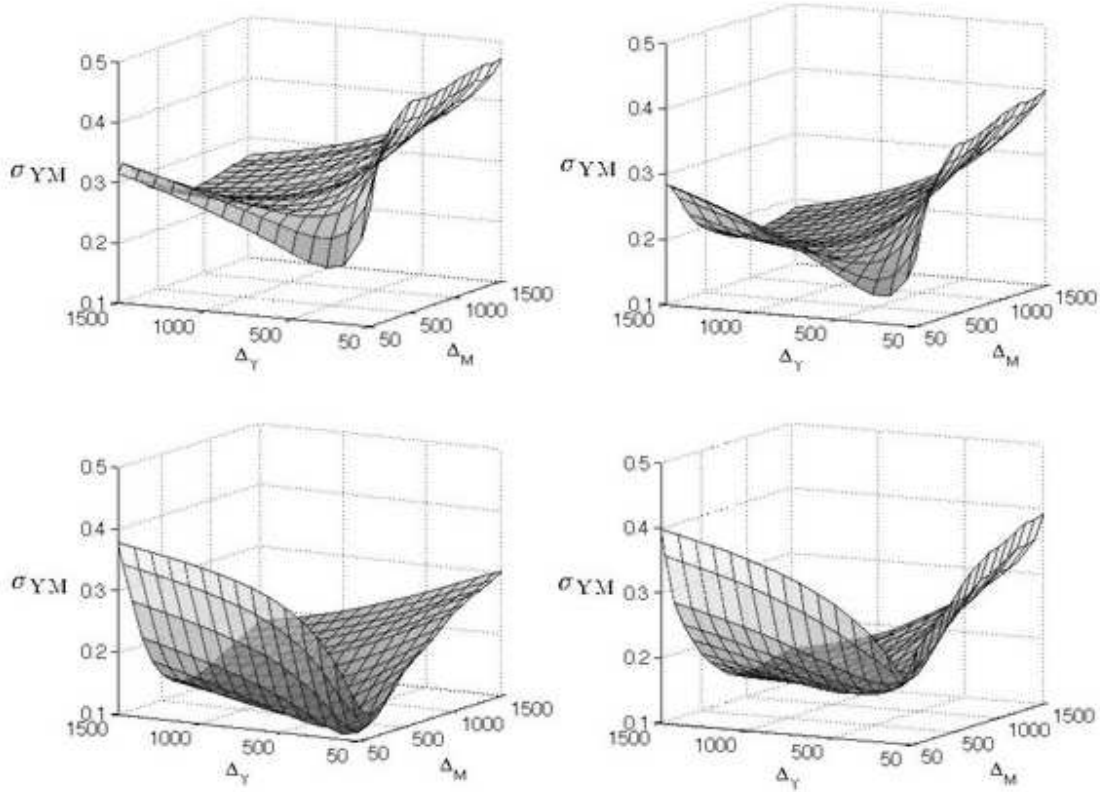}
\caption{The scatter, $\sigma_{\rm YM}$ (Eqn. \ref{eqn:scat}), around
  the mass - integrated SZE relation for cluster mass measured within
  the radius defined by the overdensity factor $\Delta_{\rm M}$ and
  the integrated SZE measured within projected radius defined by
  $\Delta_{\rm Y}$. Each panel gives the results for one of our
  cluster samples: FULL ({\it top left}), NOFB ({\it top
  right}), BASIC ({\it lower left}) and SPH ({\it lower right}).}
\label{fig:compare_models}
\end{figure*}

Figure \ref{fig:compare_models} displays the scatter $\sigma_{\rm YM}$
in the $Y-M$ relation obtained for all combinations of $\Delta_{\rm
Y}$ and $\Delta_{\rm M}$, for each of our cluster samples; FULL ({\it
top left}), NOFB ({\it top right}), BASIC ({\it bottom left}) and SPH
({\it bottom right}). We consider first the results of the BASIC
sample as it corresponds to the most simplistic case. As described in
Sec \ref{sec:gasmod}, for this sample we make the simple assumption
that the gas in clusters is isothermal and follows the dark matter
density.  Therefore, variations in the dark matter mass within
$\Delta_{\rm M}$ are also reflected in the measured flux within the
same overdensity, explaining the valley in the $\Delta_{\rm
M}$-$\Delta_{\rm Y}$ plane along $\Delta_{\rm M} = \Delta_{\rm
Y}$. This characteristic is evident in the results for all four of our
cluster samples.  The minimum scatter for the BASIC sample is near
$\Delta = 500$ for both mass and flux. This is because the temperature
for these clusters was set via the $M- T_{mw}$ scaling relation,
with the mass defined at $\Delta = 500$.

The NOFB sample exhibits similar features to the BASIC sample, the
most significant difference being the smaller rise in scatter as
$\Delta_{\rm Y}$ increases at low values of $\Delta_{\rm M}$. This is
straightforward to understand: assuming hydrostatic equilibrium results
in gas density profiles that are less sensitive to variations in the
internal mass distribution of the host dark matter halo, especially in
the inner regions, than for the clusters in the BASIC sample. The gas
density and temperature profiles are set by a combination of the total
gravitational and kinetic energy of the gas and the surface pressure
at the virial radius. We find that our recipe for star-formation
has little effect on $\sigma_{\rm YM}$.

\begin{table}
\begin{center}
\begin{tabular}{|c | c c c | c c c|}
\hline
 & & & & & & \\
$\Delta$  & \multicolumn{3}{c|}{$\langle f_{\rm gas} \rangle$ } & \multicolumn{3}{c|}{$\sigma_{\rm gas}/ \langle f_{\rm gas} \rangle$} \\
 & & & & & & \\
\hline
 & & & & & & \\
 & FULL & NOFB & SPH & FULL & NOFB & SPH \\
 & & & & & & \\
\hline
 & & & & & & \\
100 & 0.12 & 0.17 &  0.09 & 0.15 & 0.11 & 0.03 \\
 & & & & & & \\
200 & 0.11 & 0.16 &  0.09 & 0.15 & 0.11 & 0.04 \\
 & & & & & & \\
500 & 0.09 & 0.15 &  0.09 & 0.17 & 0.11 & 0.05 \\
 & & & & & & \\
1500 & 0.07 & 0.13 & 0.09 & 0.24 & 0.15 & 0.08 \\
 & & & & & & \\
\hline
\end{tabular}
\caption{Mean and fractional standard deviations of the cluster gas
fraction measured within four radii.}
\label{tab:baryon_frac}
\end{center} 
\end{table}

The FULL cluster sample -- for which we added a prescription for
energy feedback in the gas model -- resembles the NOFB sample, but
with uniformly greater scatter. Including feedback results in more
extended (or `puffed out') gas distributions, with a shallower gas
density profile in the inner regions \citep{Bode:07}. Importantly, we
have found that this substantially increased scatter is due to the
greater variance in the gas fraction within any given
overdensity. Table \ref{tab:baryon_frac} gives the mean $\langle
f_{\rm gas} \rangle$ and fractional standard deviation  $\sigma_{\rm
gas}/ \langle f_{\rm gas} \rangle$ of the gas fraction within selected
overdensities for the FULL, NOFB and SPH samples. Due to the
dependence of $f_{\rm gas}$ on cluster mass in the former, we calculate
these quantities for clusters in the mass range $10^{14} \leq M_{200}
\leq 2\times 10^{14} h^{-1} \msun$. Note that the mean values we
measure within $R_{500}$ and $R_{100} (\approx R_{\rm vir})$ are in
good agreement with the simulations of \citet{Kravtsov:05}, although
we measure greater scatter. For all three samples shown, the
fractional scatter in $f_{\rm gas}$ increases with overdensity, most
noticeably for the FULL sample.  At all radii, the percentage scatter
in $f_{\rm gas}$ for the FULL clusters is approximately 1.5 times
greater than that for the NOFB sample and more than 3 times than that
of the SPH sample. We find that $\sigma_{\rm gas}/ \langle f_{\rm gas}
\rangle$ increases when recalculated using lower mass
clusters. Repeating this exercise for gas temperature, we find the
scatter to be similar for all three models. Hence, we conclude that
the scatter in $f_{\rm gas}$ for constant cluster mass is responsible
for the increase in $\sigma_{\rm YM}$ over the three models in Table
\ref{tab:baryon_frac}.

In the bottom right panel of Figure \ref{fig:compare_models} we plot
the $\sigma_{\rm YM}$ plane for the clusters extracted from the SPH
simulation. The results are very similar to those of the BASIC cluster
sample. The scatter is least along the valley defined by $\Delta_{\rm
M} = \Delta_{\rm Y}$, rising to a maximum of $\sigma \approx 0.4$ on either
side. As discussed in Sec. \ref{sec:method}, \citet{Pearce:00}, and
\citet{Muanwong:01,Muanwong:02}, the omission of radiative cooling in
this simulation results a large central concentration of baryonic gas
(compared to models that allow for cooling and star-formation) and
thus in a ICM radial density profile that closely resembles that of
the dark matter \citep[see also,][]{daSilva:04, Ettori:06}. Both the
BASIC and SPH samples consist of clusters with very high gas density
in their centres; this is not the case for the clusters in the other
two samples.

\begin{figure}
\epsfig{file=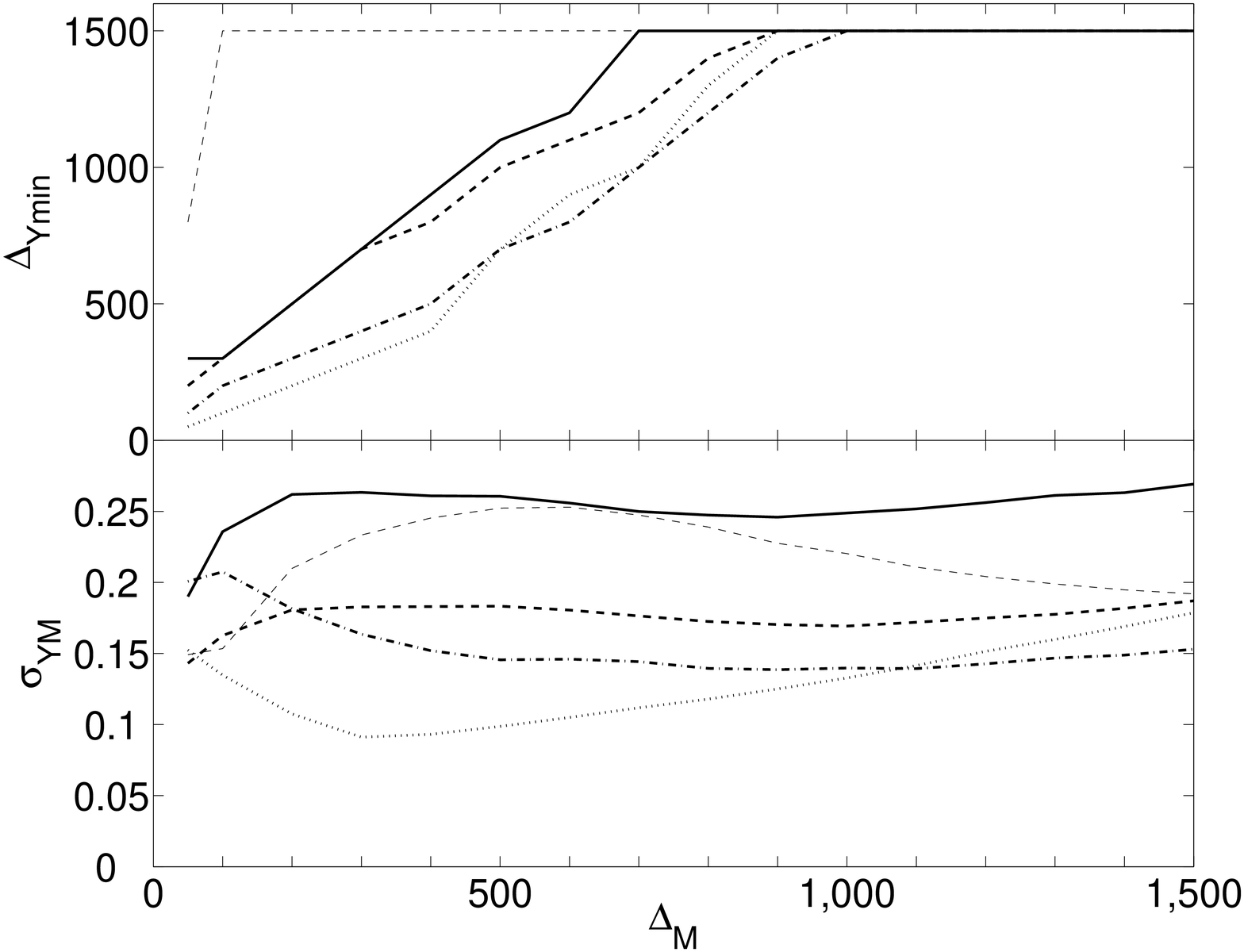,height=7cm,width=9cm}
%\plotone{BestDMDY.eps}
\caption{({\it upper}) The overdensity $\Delta_{\rm Ymin}$ for which
the scatter $\sigma_{\rm YM}$ is least for a given $\Delta_{\rm
M}$. ({\it lower}) the corresponding value of $\sigma_{\rm YM}$ for
each combination of $\Delta_{\rm M}$, $\Delta_{\rm Ymin}$. Solid,
dotted, dashed and dot-dashed lines correspond to the FULL, NOFB,
BASIC and SPH samples respectively. The light-dashed lines are the
results obtained when only clusters within a narrow concentration
range are selected from the FULL sample (see
Sec. \ref{sec:conc}). Note that where $\Delta_{\rm Ymin}$ appears to
become greater than 1500, the value of $\sigma_{\rm YM}$ in the lower
panel is calculated at this overdensity.}
\label{fig:bestDMDY}
\end{figure}

In the upper panel of Figure \ref{fig:bestDMDY} we plot the
overdensity, $\Delta_{\rm Ymin}$, at which the least-scatter measure
of the integrated SZ flux is obtained for a given $\Delta_{\rm M}$.
In the lower panel we plot the corresponding value of $\sigma_{\rm
YM}$. The solid, dotted, dashed and dot-dashed lines represent the
FULL, NOFB, BASIC and SPH cluster samples, respectively. For all four
samples $\Delta_{\rm Ymin}$ increases with $\Delta_{\rm M}$, with a
slope greater than one. For the FULL sample, the best measures of
$M_{50}$, $M_{200}$, and $M_{500}$ are $Y_{300}$, $Y_{500}$, and
$Y_{1100}$, respectively. 
For $\Delta_{\rm M} > 700$, $\Delta_{\rm Ymin}$ appears to
increase beyond the scales probed here. Performing a linear fit to
this point gives $\Delta_{\rm Ymin} = 1.9 \Delta_{\rm M} +
150$. Overall, our results show that the best measure of $M_\Delta$ is
obtained by measuring $Y$ within a smaller area than $R_{\Delta}$.

The lower panel indicates that for the FULL and NOFB models, the
scatter is least for $\Delta_{\rm M} = 50$, and is roughly constant
for $\Delta_{\rm M} > 200$.  As discussed above, much of the scatter
in these models is due to variations in the baryon fraction within
each radius.  This source of intrinsic scatter is suppressed in the
very outer regions of clusters as the baryon fraction approaches the
cosmic mean. For the SPH sample, the scatter decreases with increasing
$\Delta_{\rm M}$ to a constant value of 15\%.  For the BASIC sample,
the dip in $\sigma_{\rm YM}$ at around $\Delta_{M,Y} = 400$ (as noted
above) is clearly evident.

\subsection{Impact of halo concentration}
\label{sec:conc}
The concentration of a dark matter halo is normally defined as the
ratio of the virial radius to the NFW scale radius, $R_{vir}/r_s$ and
is thus a measure of the density in the halo core regions
\citep{Navarro:96, Navarro:97}. In their early studies, NFW postulated
that the concentration of a halo is an indicator of the mean density
of the universe at the time of its collapse \citep{Navarro:97}. Hence,
present-day clusters with a high concentration will have formed early
and remained undisturbed by major mergers, growing through gradual
accretion and accumulation of much smaller objects. Halos with a low
concentration will have had a more tumultuous recent merger history,
and may still be in the process of relaxing into dynamical equilibrium
\citep{Wechsler:02, Wechsler:06, Shaw:06, Maccio:07}.

There have now been many studies of distribution of halo concentration
and its correlation with cluster mass \citep{Jing:00, Bullock:01b,
Eke:01, Dolag:04, AvilaReese:05}. Most studies find that at a given
redshift the distribution of halo concentrations is well described by
a log-normal function of dispersion $\approx 0.22$ and mean value $<c>
\approx 5$ at z=0 \citep{Jing:00, Bullock:01b}. Halo concentration is
found to decrease as a function of increasing halo mass and is well
described by a power-law of slope in the range -0.14 to -0.1
\citep{Bullock:01b, Dolag:04, Shaw:06, Maccio:07}. However, there is
typically much scatter in this relation. One might expect to find that
variations in cluster concentration (and thus central potential) will
have a significant effect on the gas temperature and density profiles,
and thus the $Y-M$ scaling relation. We therefore fit NFW profiles and
measure the concentrations of all the halos in our samples, in order
to quantitatively measure the impact of this variation.

In Fig. 3, we recalculate $\sigma_{\rm YM}$ for the
FULL cluster sample, using only clusters with concentration $5.25 \leq
c \leq 5.75$.  The clusters in this sample are thus self-similar in
terms of their dark matter mass distribution. By comparing this figure
with the upper-left panel of Fig. \ref{fig:compare_models}, it is
clear that constraining halo concentration significantly changes the
geometry of the $\sigma_{\rm YM}$ plane. For constant $\Delta_{\rm
M}$, the scatter decreases with increasing $\Delta_{\rm Y}$. For
constant $\Delta_{\rm Y}$, $\sigma_{\rm YM}$ increases slowly as
$\Delta_{\rm M}$ decreases towards $\Delta_{\rm M} = 500$, and then
falls off rapidly in the outer regions. We have verified that similar
results are obtained for different concentration bins and our three
other cluster samples. On removing clusters from our sample for which
the NFW profile is a poor fit, the peak at $\Delta_{\rm M} = 500$
disappears, hence this feature seems to be related to clusters for
which we have a poor measure of concentration.

The light-dashed line in the upper and lower panels of Figure
\ref{fig:bestDMDY} give $\Delta_{\rm Ymin}$ and the correspond scatter
at each $\Delta_{\rm M}$ for the clusters in this concentration
range. For all but the two outermost radii, the optimal radius within
which to measure the integrated SZE appears to be less than
$R_{1500}$. (Note that where $\Delta_{\rm Ymin}$ appears to become
greater than 1500, the value of $\sigma_{\rm YM}$ in the lower panel
is calculated at this overdensity). \citet{Motl:05} demonstrated
empirically that the integrated SZE (within $R_{500}$) provides a more
accurate measure of cluster mass than the central decrement, $y$. Our
results suggest that variations in halo concentration (and thus
central potential) between clusters of the same mass may be
responsible for much of the very large scattered observed by
\citet{Motl:05} in the y-M relation. Indeed, they find that $Y_{500}$
is less sensitive to mergers than $y$; it has been previously shown
that concentration is strongly influenced by the dynamical state of a
halo \citep[e.g. ][]{Shaw:06, Maccio:07}.

\begin{figure}
\epsfig{file=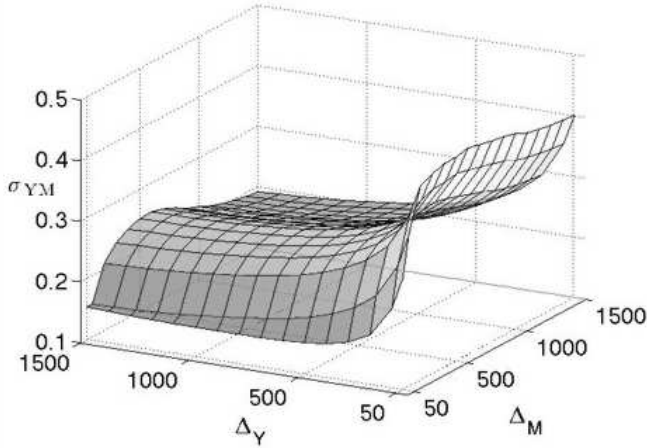,height=7cm,width=9cm}
\label{fig:const_conc}
\caption{The $\sigma_{\rm YM}$ plane calculated for the FULL cluster
sample having included only clusters with halo concentration in the range
$5.25 \leq c \leq 5.75$. It is clear that the geometry of the plane is
significantly altered compared to the results when all clusters are
included, as shown in (the upper-left panel of)
Fig. \ref{fig:compare_models}.}
\end{figure}

\subsection{Impact of Substructure}

Halo substructure is identified using the SKID algorithm of
\citet{Stadel:97}, with a smoothing length of $5\epsilon$, where
$\epsilon$ is the spline kernel force softening length of our
simulation. The minimum mass of subhalos that we resolve are $\approx
3\times 10^{11} M_\odot$, corresponding to 100 simulation
particles. For each halo, we define the substructure fraction $f_s =
M_s/M_{vir}$, where $M_s$ is the total subhalo mass. We find a mean
value $f_s = 0.084$ for our N-body halo sample, where we have included
all subhalos with centres located within the radius $R_{50}$ of the
cluster. This is in good agreement with previous studies of the
substructure content on cluster mass halos \citep{DeLucia:04, Gao:04,
Gill:04b, Shaw:07}.

We find that, on removing all clusters with $f_s > 0.084$ from the
FULL sample, $\sigma_{YM}$ is uniformly reduced by 10\%. Taking the
sample quartile with the lowest substructure fraction ($f_s < 0.050$)
we find the average decrease in $\sigma_{YM}$ is 20\% (e.g. at
$\Delta_{\rm Y} = \Delta_{\rm M} = 100$, $\sigma_{\rm YM}$ is reduced
from 0.30 to 0.25). Taking the quartile with the highest substructure
fraction ($f_s > 0.010$) increases $\sigma_{\rm YM}$ by 32\% (compared
to the result for the entire sample). Note that in all cases the
geomentry of the $\sigma_{\rm YM}$ plane does not change significantly
from that shown in the upper-left panel of
Fig. \ref{fig:compare_models}.

Overall, we find that substructure accounts for approximately a fifth
of the total amount of scatter in the $Y-M$ relation, but -- unlike the
variations in halo concentration -- does not strongly effect the
dependence of $\sigma_{\rm YM}$ on the combination of $\Delta_{\rm Y}$
and $\Delta_{\rm M}$.

\section{Impact of Projection Effects on the $Y-M$ relation}
\label{sec:projection}
Taking the standard definition of cluster mass, $M_{200}$, we have
established that the intrinsic scatter in the $Y - M_{200}$
scaling relation is least when $Y$ is measured within $R_{500}$
($\approx 0.66 {\rm R}_{200}$). Assuming that $\sigma_{\rm YM}$ does
not change significantly with redshift, for a cluster at redshift $z$
this radius translates to an optimal angular radius, $\theta_{opt}$,
within which one can obtain the least-scatter estimate of cluster mass
through an SZE measurement. However in order to make a useful estimate
of $\theta_{\rm opt}$, we must also account for the impact of the SZ
background -- the superposition of faint fore- and background clusters
along the line of sight -- when measuring the integrated SZ flux of a
cluster within some angular radius $\theta$.

Recently, \citet{Holder:07} investigated the confusion in cluster SZE
flux measurements due to the SZ background using a sample of sky maps
constructed to represent a range in $\sigma_8$ of 0.6--1. Intracluster
gas was assumed to follow the analytic model of
\citet{McCarthy:03}. They found that the mass scale below which the
$rms$ fractional errors in flux measurements become less than 20\%
occurs at just above $10^{14} h^{-1} \msun$, although this is
sensitively dependent on $\sigma_8$ and increases with decreasing
redshift. \citet{Hallman:07} explored the contribution of both gas in
low mass ($< 5 \times 10^{13} \msun$) halos and gas outside of cluster
environments on the SZE signature of resolvable clusters, using
lightcones constructed from an adiabatic hydro simulation. They find
that the integrated background SZE makes up between 4\% and 12\% of
the total cluster signal, depending on the beam size and sensitivity
of a survey, although these values are averaged over a range of
cluster masses. Both these studies concentrated on the SZ background
as the main source of scatter in SZ cluster mass measurements. In this
Section, we combine the intrinsic scatter in the $Y-M$ relation due to
internal variations between clusters ($\sigma_{\rm YM}$) with that due
to confusion with the SZ background, in order to obtain the optimal
angular radius $\theta_{opt}$ within which the integrated SZE can best
be obtained for a cluster of given mass and redshift.  Henceforth, we
fix our definition of cluster mass to the standard definition,
$M_{200}$.

For this purpose, we use the redshift range $0 \leq z \leq 1.5$
of a lightcone constructed from an N-body simulation with the same
cosmology as that used in the previous section, but with lower mass
resolution (particle mass $m_p = 1.22 \times 10^{11} \, h^{-1}
\msun$). The lower resolution was enforced due to the amount of disk
space required to store the lightcone over an octant ($\approx 5250$
square degrees) out to high redshift, and the computing time required
to analyze determine the gas distribution within each halo with the
full gas model (including star formation and feedback) within this
volume.

We proceed in the following manner. First, we measure the mean
$\langle Y_{bck} \rangle$ and standard deviation $\delta_{bck}$ of
the background integrated SZE within a circular aperture of radius $\theta$
centred around each cluster in a sample of images generated from the
lightcone. We do this for a range of angular radii, varying $\theta$
from 0.5 to 15 arcminutes. We then obtain the total scatter in the
integrated SZE measured within $\theta$ for a cluster of mass
$M_{200}$ at a redshift $z$ by combining in quadrature the scatter due
to variations in the background flux with that due to variations in
the internal properties of clusters, as measured in
Sec. \ref{sec:results}.

To measure $\langle Y_{bck} \rangle$ and $\delta_{bck}$ we randomly
select a sample of 100 clusters, each with mass $M_{200} \geq 10^{14}
h^{-1} \msun$, over a redshift range $0.2 \leq z \leq 1$ from the
lightcone.  For each cluster we generate two images, one including and
one omitting all the foreground and background clusters within a one
degree radius around the cluster.  We then remove the latter from the
former, leaving just an image of the interlopers. We wish to
investigate the impact of the SZ background due to {\it dim} clusters
along the line of sight, therefore we also remove all clusters with an
SZ signal to noise $\geq 5$ from the images, where $Y$ is measured
within $\theta(R_{200}, z)$ and we assume an instrument noise of
$10\mu$K. These clusters are then removed in the following
manner. First, the central pixel of the interloper is identified and
the flux $y_{max}$ recorded. Next, a beta model of the form
\begin{equation}
y(\theta) = y_{max}(1 + \theta^2/\theta_c^2)^{-0.5}
\end{equation}
is deducted from the image, where $\theta_c \approx R_{vir}/20D_a$, and
$R_{vir}$ and the angular diameter distance, $D_a$, are taken directly
from our lightcone halo catalog (we remove the cluster signal out
to the angular size corresponding to the virial radius of each
cluster).

Note that we do not add the primary CMB signal or other sources of
`noise' -- radio point sources, instrumental noise, galactic dust
emission and atmospheric emission \citep[see, for
example,][]{Sehgal:07} -- into these images. In a subsequent paper
(Shaw \& Holder, {\it in preparation}) we will investigate in more
depth the accuracy with which matched-filtering schemes
\citep[e.g.][]{Melin:06} can measure clusters masses, incorporating
some of these effects.  Here we focus solely on sources of scatter in
the $Y-M$ relation that are due to variations in the internal properties
and spatial distributions of clusters, and thus cannot be suppressed.

\begin{figure}
\epsfig{file=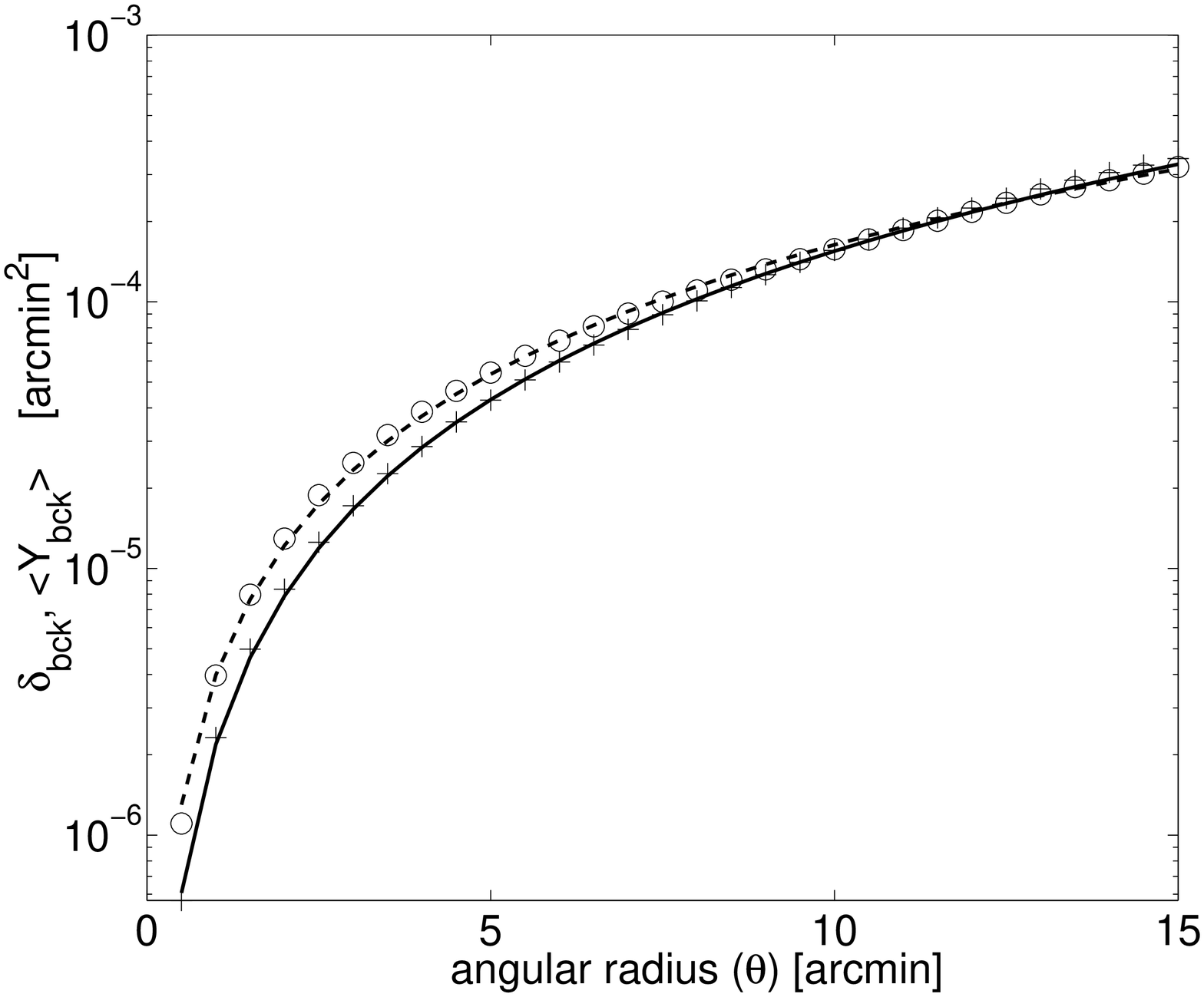,height=7cm,width=9cm}
%\plotone{projected.eps}
\caption{The mean ($\langle {\rm Y_{bck}}\rangle$, crosses) and
standard deviation ($\delta_{\rm bck}$ , circles) of the background SZE
flux measured within an angular region of radius $\theta$ centred
around each bright cluster in our sample of images, due to dim,
unresolvable clusters (see text). Solid and dashed lines give
power-law fits to the results, as given in Eqns. \ref{eqn:Ybck} and
\ref{eqn:deltabck}.}
\label{fig:backgroundflux}
\end{figure}

We measure the mean and standard deviation of the background
integrated SZE at each value of $\theta$ over the selected cluster
sample. The results are displayed in
Fig. \ref{fig:backgroundflux}. The crosses represent the mean
integrated flux $\langle {\rm Y_{bck}}\rangle$ within $\theta$. The
circles represent the standard deviation $\delta_{\rm bck}$ around the
mean for each $\theta$. We find that both are well described by a
power-law, with 
\begin{equation}
\langle {\rm Y_{bck}} \rangle = 10^{-5.66}\theta^{1.85}
\label{eqn:Ybck}
\end{equation}
and 
\begin{equation}
\delta_{\rm bck} = 10^{-5.4}\theta^{1.61} \;,
\label{eqn:deltabck}
\end{equation}
plotted as solid and dashed lines, respectively. We note that both
$\langle {\rm Y_{bck}}\rangle$ and $\delta_{\rm bck}$ are dependent on
the matter power spectrum and thus on the values of $\Omega_M$ and
$\sigma_8$ \citep[as demonstrated by ][]{Holder:07} and therefore the
values measured here are only relevant for the cosmology assumed by
our simulation.

Using Equations \ref{eqn:Ybck} and \ref{eqn:deltabck}, and the fitted
values for the $Y_{\Delta}-M_{200}$ scaling relations measured in
Section. \ref{sec:results}, we are now able to calculate the mean
total flux that would be measured within an angular radius
$\theta(\Delta_{\rm Y}, z)$ for a cluster of mass $M_{200}$ at redshift
$z$:
\begin{equation}
Y_{tot}(M_{200}, \theta) = \langle Y_{bck}(\theta) \rangle + Y_{\rm clus}(M_{200},
\theta, z) \;,
\end{equation}
where
\begin{equation}
Y_{\rm clus}(M_{200}, \theta, z) =
\frac{E(z)^{3/2}}{d^2_A(z)}10^{A(\Delta)}\left (\frac{M_{200}}{10^{14}
h^{-1} \msun}\right)^{\alpha(\Delta)}
\end{equation}
and $\theta = \theta(\Delta, z)$ is the angular radius corresponding
to a cluster with radius $R_\Delta$ at angular diameter distance
$d_A(z)$ (note the units of $Y_{\rm clus}$ are arcmin$^2$, see
Eqn. \ref{eqn:integratedy}). $A(\Delta)$ and $\alpha(\Delta)$ are the
corresponding normalisation index and slope obtained for the
$Y_\Delta-M_{200}$ relation as measured in Section
\ref{sec:results}.  The radius at a given $\Delta$ is calculated at
redshift $z$ by relating $R_{\Delta}$ to $R_{200}$ through the NFW
profile. We use the mass-concentration fitting formula of
\cite{Dolag:04} to calculate the mean concentration for a cluster of
mass $M_{200}$ at each cluster redshift. We assume that the slope and
normalisation of the $Y-M$ relation are independent of redshift (having
accounted for the hubble scaling, $E(z)$). \citet{Nagai:06} have found
this to be the case for a simulated sample of clusters encompassing a
redshift range $0 \geq z \geq 2$.

We can calculate the total scatter in $Y_{tot}(\theta, M, z)$ by
combining the scatter in the mean background flux with the intrinsic
scatter in the $Y-M$ relation $\sigma_{\rm YM}(\Delta_{\rm Y})$;
\begin{equation}
\sigma_{tot}(\theta, M, z) = \frac{(\delta_{bck(\theta)}^2 +
\delta_{clus}^2)^{\frac{1}{2}}}{Y_{\rm tot}} \;,
\label{eqn:totscat}
\end{equation}
where
\begin{equation}
\delta_{clus} = Y_{\rm clus}\sigma_{\rm YM}(\Delta_{\rm Y}) \;,
\label{eqn:delta_clus}
\end{equation}
thus converting the {\it fractional} scatter $\sigma_{\rm YM} \approx
\delta_{\rm clus}/Y_{\rm clus}$, to an absolute value.

\begin{figure}
%\plotone{projected2.eps}
\epsfig{file=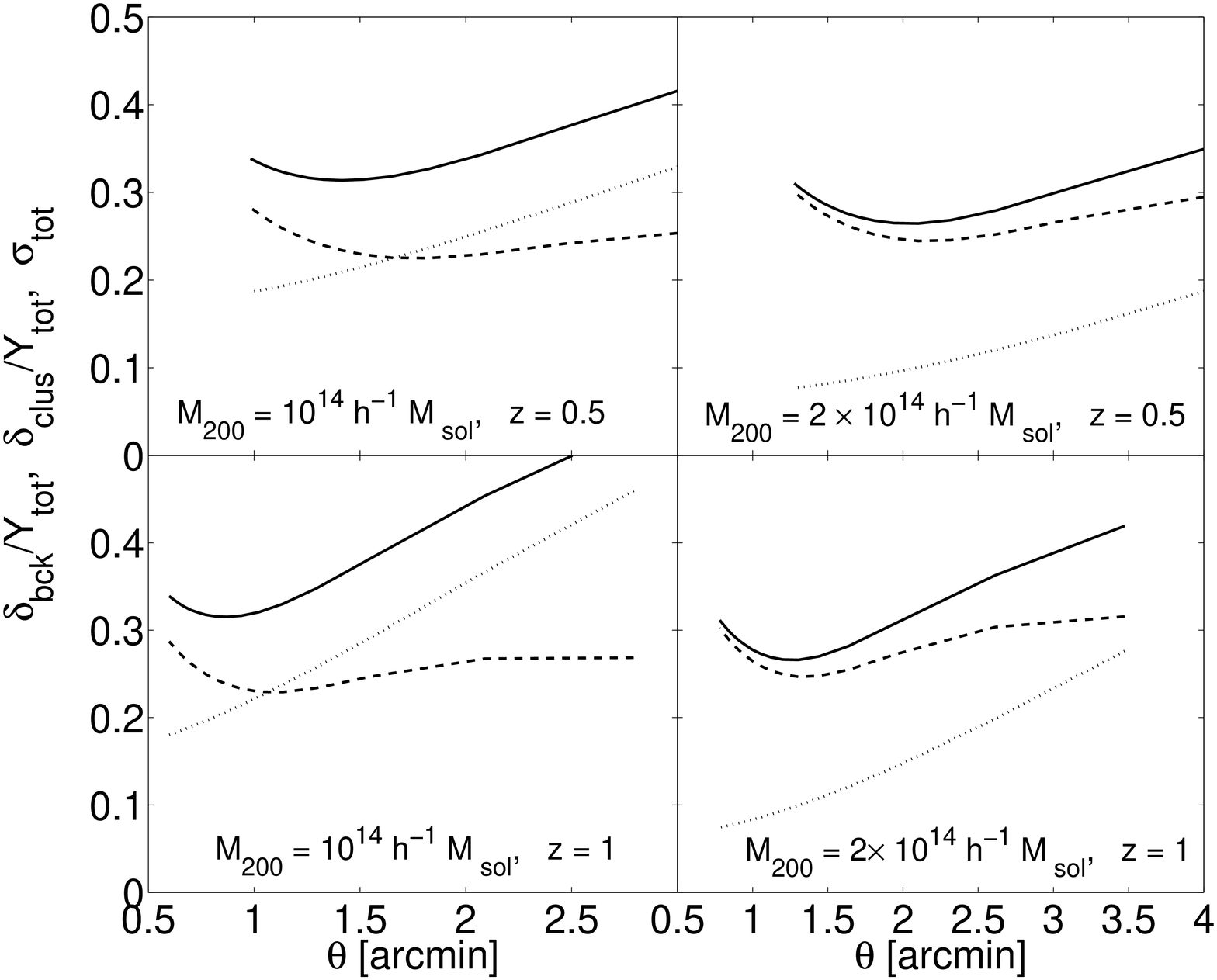,height=7cm,width=9cm}
\caption{({\it solid lines)} The total scatter $\sigma_{\rm tot}$
(Eqn. \ref{eqn:totscat}) in the $Y-M$ relation for $Y$ measured within
angular radius $\theta$ for a $10^{14} h^{-1} \msun$ ({\it left
panels}) and $2\times 10^{14} h^{-1} \msun$ ({\it right}) cluster at
redshifts 0.5 ({\it upper panels}) and 1 ({\it lower}). The dashed
lines give the fractional variation in the intrinsic scatter
($\delta_{\rm clus}/Y_{\rm tot}$, Eqn. \ref{eqn:delta_clus}), the
dotted lines give the fractional variation in the SZE background
($\delta_{\rm bck}/Y_{\rm tot}$), within $\theta$ for each cluster
mass.}
\label{fig:scat_tot}
\end{figure}

Figure \ref{fig:scat_tot} demonstrates the dependence of
$\sigma_{tot}$ (solid lines) on angular radius for clusters of mass
$10^{14} h^{-1} \msun$ and $2\times 10^{14} h^{-1} \msun$, at
redshifts $z = 0.5$ and $1$.  Also plotted are $\delta_{\rm
clus}/Y_{\rm tot}$ (dashed) and $\delta_{\rm bck}/Y_{\rm tot}$
(dotted). The minimum of the dashed line corresponds to angular radius
subtending $R_{500}$ for each cluster mass and redshift. For a
$10^{14} h^{-1} \msun$ cluster (left panels), including the SZE
background moves the optimal value of $\theta$ to a lower angular
radii than that predicted by the intrinsic cluster scatter
$\delta_{\rm clus}$ alone. At z = 0.5, this decrease is 0.4
arcminutes, and at z = 1 it is 0.25 arcminutes. At higher redshifts,
the cluster subtends a smaller angular region and thus the impact of
the background flux is lessened. This is also the case as cluster mass
is increased. The right panels of Figure \ref{fig:scat_tot}
demonstrate that, for a $2\times 10^{14} h^{-1} \msun$ cluster, the
background flux is small compared to the cluster signal, and therefore
$\delta_{\rm bck}$ does not change $\theta$ significantly. Hence, for
clusters of mass $M_{200} \geq 2\times 10^{14} h^{-1} \msun$
variations in the SZE background are negligible compared to the
intrinsic scatter in $Y$ due to variations in the internal properties of
clusters.

\begin{figure}
%\plotone{projected2.eps}
\epsfig{file=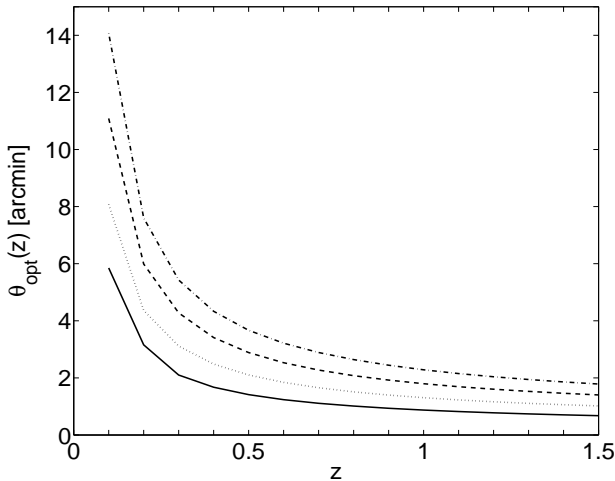,height=7cm,width=9cm}
\caption{The optimal angle $\theta_{\rm opt}$ within which SZE flux
can be measured for a cluster of mass $10^{14}$ ({\it solid
line}), $2\times 10^{14}$ ({\it dotted}), $5\times 10^{14}$
({\it dashed}) and $10^{15}$ ({\it dot-dashed}) $h^{-1}
\msun$ to give the last scatter (intrinsic plus projected) in the $Y-M$
relation cluster at redshift $z$.}
\label{fig:best_ang}
\end{figure}

Finally, in Fig. \ref{fig:best_ang} we plot the optimal angle
$\theta_{\rm opt}(z)$ within which the integrated SZ signal provides
the least scatter measure of $M_{200}$ as a function of redshift. The
solid, dotted, dashed and dot-dashed lines correspond to 1,2,5 and 10
$\times 10^{14} h^{-1} \msun$ clusters respectively. Thus $\theta_{\rm
opt}(z=0.5)$ varies between 1.5 and 4 arcminutes as mass increases.  For
$z>1$, all clusters have an optimal angular radius less than 2
arcminutes.

\section{Discussion and Conclusions}
\label{sec:discussion}

The Sunyaev-Zel'dovich effect is one of the most promising means for
precise determination of cosmological parameters because the mapping
between the integrated SZ flux and cluster mass is expected to have
very low intrinsic scatter.  Thus there is a need for concerted
investigation into the origins and character of the scatter that is
present in this relation, and how the scatter may be reduced.

In this paper, we have presented a detailed analysis of the intrinsic
scatter in the integrated SZE--cluster mass ($Y-M$) relation, using
large simulated cluster samples generated from a semi-analytic model
of intracluster gas applied to the lightcone output of a high
resolution N-body simulation, and also clusters extracted directly
from an adiabatic SPH simulation. In particular, the main aims of this
study were to investigate the impact on this relation of:
\begin{itemize}
\item the choice of how cluster mass is defined and the region within
which the integrated SZE is measured; i.e. the impact of measuring
these quantities within different radii $R_{\Delta}$, where $\Delta$
is the overdensity relative to the critical density,
\item incorporating energy feedback due to supernovae and AGN outflows
into the cluster gas model,
\item variations in host halo concentration and substructure populations,
\item the error in SZ flux measurements due to confusion with
unresolvable (i.e. low Y) clusters along the line of sight.
\end{itemize}

To assess the impact of the definition of $Y$ and $M$, we measure both
quantities within radii corresponding to a range of overdensities
between $\Delta = 50$ and $1500$. We measure the slope, normalisation
and scatter in the ${\rm Y_\Delta-M_\Delta}$ relation using {\it all}
combinations of $\Delta_Y$ and $\Delta_M$ (not just $\Delta_Y =
\Delta_M$). The effect of energy feedback on the scatter is
investigated by generating three realisations of our gas model,
respectively including and omitting energy feedback and a toy model in
which the gas is isothermal and follows the dark matter density. Each
sample contained 1267 clusters. For comparison, we also take
a sample of 212 clusters from the output of an adiabatic simulation
\citep{Muanwong:01}.  For each sample, we also measure the
concentration and fraction of mass contained in substructure of the
host dark matter halos. 

Finally, we address the issue of the SZE background using the full
lightcone output of our N-body simulation. We measure the SZE
background due to low mass systems within a range of solid angles,
combining the scatter in the background flux with the intrinsic
scatter in the $Y-M$ relation (within a given angular radius) to
determine the optimal angular size within which the scatter in the
measured $Y$ for a cluster of given mass and redshift is least. Our
results are summarized below.

\begin{enumerate}
\item Scatter in the intrinsic $Y-M$ relation is least when cluster mass
is defined within as large a radius as possible (to a maximum of
$R_{50}$ in this study) and the integrated SZE is measured in the
range $R_{500} \leq R_{\Delta} \leq R_{300}$. For $M_{50}$ the scatter
is least for $Y_{300}$. For $M_{200}$, the scatter is least for
$Y_{500}$. In general, the least scatter measure of $Y$ is always
obtained within a smaller radii than that at which the mass is
defined. We find that the best overdensity within which to measure Y
is given by $\Delta_{\rm Y} = 1.9 \Delta_M + 150$.
\item Inclusion of energy feedback in our gas model significantly
increases the intrinsic scatter in the $Y-M$ relation for all
combinations of $\Delta_{\rm M}$ and $\Delta_{\rm Y}$. We find that
this is due to significantly larger variations in the gas mass
fraction within any radius than in our other cluster samples.
\item Variations in host halo concentration/central potential (for
clusters of the same mass), and the resulting impact on the cluster
gas distribution, provide a reason as to why the integrated SZE
provides a tighter proxy for cluster mass than the central decrement.
Constraining halo concentration (i.e. selecting only clusters within a
narrow concentration range) changes the $Y-M$ relation for high $
\Delta_{\rm Y}$ so that the least scatter measure of $M_\Delta$ is
obtained by measuring the integrated SZE within a radius $R \leq
R_{1500}$, the minimum radius probed in this study.
\item Substructure increases scatter uniformly for all combinations of
$\Delta_{\rm M}$ and $\Delta_{\rm Y}$. Removing clusters with large
amounts of substructure reduces scatter in the $Y-M$ relation by up to
20\%.
\item The mean integrated SZE around a cluster due to low-mass
foreground and background clusters along the line of sight scales with
angular radius $\langle Y_{bck} \rangle = 10^{-5.66} \theta^{1.85}$, and the
standard deviation as $\delta_{bck} = 10^{-5.4}\theta^{1.61}$.
\item For cluster of mass $M_{200} \geq 2 \times 10^{14} h^{-1}
\msun$, the optimal angular radius within which one should measure the
SZE so as to obtain the tightest scaling with cluster mass ($M_{200}$)
is just the angle subtending $R_{500}$ at the cluster redshift (the
radius at which intrinsic scatter due to variations in internal
cluster properties is least). Below this mass, the magnitude and
variance of the SZE background relative to the cluster signal reduces
the optimal angular radius (e.g by 0.4 arcminutes at z = 0.5 for a
$10^{14} h^{-1}\msun$ cluster).  In addition, we have provided a chart
(Fig. \ref{fig:best_ang}) giving the optimal angular size for
measuring clusters masses using the SZE for a range of masses and
redshifts.
\end{enumerate}

One important result of this study is that we measure there to be {\it
at least} 20\% intrinsic scatter in the $Y-M$ relation due variations in
the internal properties of clusters alone. When mass and SZ flux are
measured at $R_{200}$, the scatter is just above 25\%. This is higher
than the 10-15\% scatter that has been quoted by previous authors
using small samples of clusters extracted from highly sophisticated
simulations including complex gas physics \citep[e.g.][]{Nagai:06} or
larger samples of less-well resolved clusters \citep{daSilva:04,
Motl:05, Hallman:07}. As noted above, we have found that much of the
scatter we measure is due to the variations in the gas mass fraction
between clusters of similar mass, an effect that is greatly enhanced
by the inclusion of energy feedback into our models. However,
observational measures of the gas fraction from X-ray studies of
bright clusters certainly seem to suggest that there is indeed much
variation in this quantity between clusters \citep{Vikhlinin:06,
Zhang:07}. Nevertheless, we intend to perform a rigorous comparison
between the predictions of our gas model and the results of
hydrodynamical simulations within the near future.

We have shown that a significant fraction of the scatter in the $Y-M$
scaling relation is due to variations in the host halo properties,
specifically in concentration and substructure populations. Our
results suggest that, if all halos at a given mass had the same merger
history, then the central SZE decrement might provide the best measure
of halo mass. However, the variations in cluster structure due to
mergers greatly and preferentially increase the scatter in Y when
measured only within the central regions ($R_{1500}$), hence it becomes
necessary to integrate to larger radii. We have shown that there is a
limit to how far out one should go, however, before projection effects
due to dim, unresolvable clusters begin to significant effect SZE flux
measurements.

We have not investigated the impact of morphology and environment in
this study, which may also play significant roles. We leave this to
future work. Furthermore, our gas model is not able to accurately
account for some of the complex physical processes such as shock
heating during mergers and cooling in cluster cores, which will
certainly introduce additional scatter in scaling relations
\citep{McCarthy:04, OHara:06, Poole:07}.

Finally, several studies have demonstrated there to be significant
errors in extracting the correct value of $Y$ from synthetic sky maps
due to CMB confusion, instrument noise, the effect of galactic dust
emission, diffuse gas outside of clusters and radio point sources, and
systematic effects in the cluster identification algorithm utilized
\citep{Melin:06, Hallman:07, Juin:07, Schafer:06, Schafer:06b,
Pires:06, Sehgal:07}. Combined with the intrinsic scatter in the $Y-M$
as measured here, it is clear that there is still much work to be done
to enable accurate estimation of cluster masses and thus achieve the
tight cosmological constraints that are envisaged from SZ cluster
surveys.

\section{ACKNOWLEDGMENTS}
This work supported by the Natural Sciences and Engineering Research
Council (Canada) through the Discovery Grant Awards to GPH. GPH would
also like to acknowledge support from the Canadian Institute for
Advance Research and the Canada Research Chairs Program. This research
was facilitated by allocations of time on the COSMOS supercomputer at
DAMTP in Cambridge, a UK-CCC facility supported by HEFCE and PPARC,
and advanced computing resources from the Pittsburgh Supercomputing
Center and the National Center for Supercomputing Applications.  In
addition, computational facilities at Princeton supported by NSF grant
AST-0216105 were used, as well as high performance computational
facilities supported by Princeton University under the auspices of the
Princeton Institute for Computational Science and Engineering
(PICSciE) and the Office of Information Technology (OIT). We would
also like to thank A. Evrard and J.P. Ostriker for helpful discussions.

%\def\apj{Ap.\ J.}  
%\bibliography{biblist.bib}

\begin{thebibliography}{85}
\expandafter\ifx\csname natexlab\endcsname\relax\def\natexlab#1{#1}\fi

\bibitem[{{Ascasibar} {et~al.}(2006){Ascasibar}, {Sevilla}, {Yepes},
  {M{\"u}ller}, \& {Gottl{\"o}ber}}]{Ascasibar:06}
{Ascasibar}, Y., {Sevilla}, R., {Yepes}, G., {M{\"u}ller}, V., \&
  {Gottl{\"o}ber}, S. 2006, \mnras, 371, 193

\bibitem[{{Avila-Reese} {et~al.}(2005){Avila-Reese}, {Col{\'{\i}}n},
  {Gottl{\"o}ber}, {Firmani}, \& {Maulbetsch}}]{AvilaReese:05}
{Avila-Reese}, V., {Col{\'{\i}}n}, P., {Gottl{\"o}ber}, S., {Firmani}, C., \&
  {Maulbetsch}, C. 2005, \apj, 634, 51

\bibitem[{Bahcall \& Fan(1998)}]{bahcall:98}
Bahcall, N. \& Fan, X. 1998, \apj, 504, 1

\bibitem[{{Balogh} {et~al.}(1999){Balogh}, {Babul}, \& {Patton}}]{Balogh:99}
{Balogh}, M.~L., {Babul}, A., \& {Patton}, D.~R. 1999, \mnras, 307, 463

\bibitem[{Barbosa {et~al.}(1996)Barbosa, Bartlett, Blanchard, \&
  Oukbir}]{barbosa:96}
Barbosa, D., Bartlett, J., Blanchard, A., \& Oukbir, J. 1996, \aap, 314, 13

\bibitem[{{Benson} {et~al.}(2004){Benson}, {Church}, {Ade}, {Bock}, {Ganga},
  {Henson}, \& {Thompson}}]{Benson:04}
{Benson}, B.~A., {Church}, S.~E., {Ade}, P.~A.~R., {Bock}, J.~J., {Ganga},
  K.~M., {Henson}, C.~N., \& {Thompson}, K.~L. 2004, \apj, 617, 829

\bibitem[{Birkinshaw(1999)}]{birkinshaw:99}
Birkinshaw, M. 1999, Physics Reports, 310, 97

\bibitem[{{Bode} {et~al.}(2007){Bode}, {Ostriker}, {Weller}, \&
  {Shaw}}]{Bode:07}
{Bode}, P., {Ostriker}, J.~P., {Weller}, J., \& {Shaw}, L. 2007, \apj, 663, 139

\bibitem[{{Borgani} {et~al.}(2005){Borgani}, {Finoguenov}, {Kay}, {Ponman},
  {Springel}, {Tozzi}, \& {Voit}}]{Borgani:05}
{Borgani}, S., {Finoguenov}, A., {Kay}, S.~T., {Ponman}, T.~J., {Springel}, V.,
  {Tozzi}, P., \& {Voit}, G.~M. 2005, \mnras, 361, 233

\bibitem[{{Bryan} \& {Norman}(1998)}]{Bryan:98}
{Bryan}, G.~L. \& {Norman}, M.~L. 1998, \apj, 495, 80

\bibitem[{{Bullock} {et~al.}(2001){Bullock}, {Kolatt}, {Sigad}, {Somerville},
  {Kravtsov}, {Klypin}, {Primack}, \& {Dekel}}]{Bullock:01b}
{Bullock}, J.~S., {Kolatt}, T.~S., {Sigad}, Y., {Somerville}, R.~S.,
  {Kravtsov}, A.~V., {Klypin}, A.~A., {Primack}, J.~R., \& {Dekel}, A. 2001,
  \mnras, 321, 559

\bibitem[{Carlstrom {et~al.}(2002)Carlstrom, , Holder, \& Reese}]{carlstrom:02}
Carlstrom, J.~E., , Holder, G.~P., \& Reese, E.~D. 2002, \araa, 40, 643

\bibitem[{{Cole} \& {Lacey}(1996)}]{Cole:96}
{Cole}, S. \& {Lacey}, C. 1996, \mnras, 281, 716

\bibitem[{{Couchman} {et~al.}(1995){Couchman}, {Thomas}, \&
  {Pearce}}]{Couchman:95}
{Couchman}, H.~M.~P., {Thomas}, P.~A., \& {Pearce}, F.~R. 1995, \apj, 452, 797

\bibitem[{{da Silva} {et~al.}(2004){da Silva}, {Kay}, {Liddle}, \&
  {Thomas}}]{daSilva:04}
{da Silva}, A.~C., {Kay}, S.~T., {Liddle}, A.~R., \& {Thomas}, P.~A. 2004,
  \mnras, 348, 1401

\bibitem[{{De Lucia} {et~al.}(2004){De Lucia}, {Kauffmann}, {Springel},
  {White}, {Lanzoni}, {Stoehr}, {Tormen}, \& {Yoshida}}]{DeLucia:04}
{De Lucia}, G., {Kauffmann}, G., {Springel}, V., {White}, S.~D.~M., {Lanzoni},
  B., {Stoehr}, F., {Tormen}, G., \& {Yoshida}, N. 2004, \mnras, 348, 333

\bibitem[{{Dolag} {et~al.}(2004){Dolag}, {Bartelmann}, {Perrotta},
  {Baccigalupi}, {Moscardini}, {Meneghetti}, \& {Tormen}}]{Dolag:04}
{Dolag}, K., {Bartelmann}, M., {Perrotta}, F., {Baccigalupi}, C., {Moscardini},
  L., {Meneghetti}, M., \& {Tormen}, G. 2004, \aap, 416, 853

\bibitem[{{Eke} {et~al.}(1996){Eke}, {Cole}, \& {Frenk}}]{Eke:96}
{Eke}, V.~R., {Cole}, S., \& {Frenk}, C.~S. 1996, \mnras, 282, 263

\bibitem[{{Eke} {et~al.}(1998{\natexlab{a}}){Eke}, {Cole}, {Frenk}, \& {Patrick
  Henry}}]{Eke:98b}
{Eke}, V.~R., {Cole}, S., {Frenk}, C.~S., \& {Patrick Henry}, J.
  1998{\natexlab{a}}, \mnras, 298, 1145

\bibitem[{{Eke} {et~al.}(1998{\natexlab{b}}){Eke}, {Navarro}, \&
  {Frenk}}]{Eke:98}
{Eke}, V.~R., {Navarro}, J.~F., \& {Frenk}, C.~S. 1998{\natexlab{b}}, \apj,
  503, 569

\bibitem[{{Eke} {et~al.}(2001){Eke}, {Navarro}, \& {Steinmetz}}]{Eke:01}
{Eke}, V.~R., {Navarro}, J.~F., \& {Steinmetz}, M. 2001, \apj, 554, 114

\bibitem[{{Ettori} {et~al.}(2006){Ettori}, {Dolag}, {Borgani}, \&
  {Murante}}]{Ettori:06}
{Ettori}, S., {Dolag}, K., {Borgani}, S., \& {Murante}, G. 2006, \mnras, 365,
  1021

\bibitem[{{Evrard} {et~al.}(2007){Evrard}, {Bialek}, {Busha}, {White}, {Habib},
  {Heitmann}, {Warren}, {Rasia}, {Tormen}, {Moscardini}, {Power}, {Jenkins},
  {Gao}, {Frenk}, {Springel}, {White}, \& {Diemand}}]{Evrard:07}
{Evrard}, A.~E., {Bialek}, J., {Busha}, M., {White}, M., {Habib}, S.,
  {Heitmann}, K., {Warren}, M., {Rasia}, E., {Tormen}, G., {Moscardini}, L.,
  {Power}, C., {Jenkins}, A.~R., {Gao}, L., {Frenk}, C.~S., {Springel}, V.,
  {White}, S.~D.~M., \& {Diemand}, J. 2007, ArXiv Astrophysics e-prints

\bibitem[{{Evrard} \& {Henry}(1991)}]{Evrard:91}
{Evrard}, A.~E. \& {Henry}, J.~P. 1991, \apj, 383, 95

\bibitem[{{Gao} {et~al.}(2004){Gao}, {White}, {Jenkins}, {Stoehr}, \&
  {Springel}}]{Gao:04}
{Gao}, L., {White}, S.~D.~M., {Jenkins}, A., {Stoehr}, F., \& {Springel}, V.
  2004, \mnras, 355, 819

\bibitem[{{Gill} {et~al.}(2004){Gill}, {Knebe}, {Gibson}, \&
  {Dopita}}]{Gill:04b}
{Gill}, S.~P.~D., {Knebe}, A., {Gibson}, B.~K., \& {Dopita}, M.~A. 2004,
  \mnras, 351, 410

\bibitem[{Haiman {et~al.}(2001)Haiman, Mohr, \& Holder}]{haiman:00}
Haiman, Z., Mohr, J.~J., \& Holder, G.~P. 2001, \apj, 553, 545

\bibitem[{{Hallman} {et~al.}(2007){Hallman}, {O'Shea}, {Burns}, {Norman},
  {Harkness}, \& {Wagner}}]{Hallman:07}
{Hallman}, E.~J., {O'Shea}, B.~W., {Burns}, J.~O., {Norman}, M.~L., {Harkness},
  R., \& {Wagner}, R. 2007, ArXiv e-prints, 704

\bibitem[{{Holder} {et~al.}(2007){Holder}, {McCarthy}, \& {Babul}}]{Holder:07}
{Holder}, G., {McCarthy}, I.~G., \& {Babul}, A. 2007, ArXiv Astrophysics
  e-prints

\bibitem[{{Hu}(2003)}]{hu:03}
{Hu}, W. 2003, \prd, 67, 081304

\bibitem[{{Jing}(2000)}]{Jing:00}
{Jing}, Y.~P. 2000, \apj, 535, 30

\bibitem[{{Juin} {et~al.}(2007){Juin}, {Yvon}, {R{\'e}fr{\'e}gier}, \&
  {Y{\`e}che}}]{Juin:07}
{Juin}, J.~B., {Yvon}, D., {R{\'e}fr{\'e}gier}, A., \& {Y{\`e}che}, C. 2007,
  \aap, 465, 57

\bibitem[{{Kaiser}(1991)}]{Kaiser:91}
{Kaiser}, N. 1991, \apj, 383, 104

\bibitem[{{Kosowsky}(2003)}]{Kosowsky:03}
{Kosowsky}, A. 2003, New Astronomy Review, 47, 939

\bibitem[{{Kravtsov} {et~al.}(2002){Kravtsov}, {Klypin}, \&
  {Hoffman}}]{Kravtsov:02}
{Kravtsov}, A.~V., {Klypin}, A., \& {Hoffman}, Y. 2002, \apj, 571, 563

\bibitem[{{Kravtsov} {et~al.}(2005){Kravtsov}, {Nagai}, \&
  {Vikhlinin}}]{Kravtsov:05}
{Kravtsov}, A.~V., {Nagai}, D., \& {Vikhlinin}, A.~A. 2005, \apj, 625, 588

\bibitem[{{Lacey} \& {Cole}(1993)}]{Lacey:93}
{Lacey}, C. \& {Cole}, S. 1993, \mnras, 262, 627

\bibitem[{{Lahav} {et~al.}(1991){Lahav}, {Lilje}, {Primack}, \&
  {Rees}}]{Lahav:91}
{Lahav}, O., {Lilje}, P.~B., {Primack}, J.~R., \& {Rees}, M.~J. 1991, \mnras,
  251, 128

\bibitem[{{Lima} \& {Hu}(2004)}]{lima:04}
{Lima}, M. \& {Hu}, W. 2004, \prd, 70, 043504

\bibitem[{{Lima} \& {Hu}(2005)}]{lima:05}
---. 2005, \prd, 72, 043006

\bibitem[{{Lin} {et~al.}(2003){Lin}, {Mohr}, \& {Stanford}}]{Lin:03}
{Lin}, Y.-T., {Mohr}, J.~J., \& {Stanford}, S.~A. 2003, \apj, 591, 749

\bibitem[{{Macci{\`o}} {et~al.}(2007){Macci{\`o}}, {Dutton}, {van den Bosch},
  {Moore}, {Potter}, \& {Stadel}}]{Maccio:07}
{Macci{\`o}}, A.~V., {Dutton}, A.~A., {van den Bosch}, F.~C., {Moore}, B.,
  {Potter}, D., \& {Stadel}, J. 2007, \mnras, 378, 55

\bibitem[{{Majumdar} \& {Mohr}(2003)}]{majumdar:03}
{Majumdar}, S. \& {Mohr}, J.~J. 2003, \apj, 585, 603

\bibitem[{{McCarthy} {et~al.}(2003){McCarthy}, {Babul}, {Holder}, \&
  {Balogh}}]{McCarthy:03}
{McCarthy}, I.~G., {Babul}, A., {Holder}, G.~P., \& {Balogh}, M.~L. 2003, \apj,
  591, 515

\bibitem[{{McCarthy} {et~al.}(2004){McCarthy}, {Balogh}, {Babul}, {Poole}, \&
  {Horner}}]{McCarthy:04}
{McCarthy}, I.~G., {Balogh}, M.~L., {Babul}, A., {Poole}, G.~B., \& {Horner},
  D.~J. 2004, \apj, 613, 811

\bibitem[{{Melin} {et~al.}(2006){Melin}, {Bartlett}, \&
  {Delabrouille}}]{Melin:06}
{Melin}, J.-B., {Bartlett}, J.~G., \& {Delabrouille}, J. 2006, \aap, 459, 341

\bibitem[{{Morandi} {et~al.}(2007){Morandi}, {Ettori}, \&
  {Moscardini}}]{Morandi:07}
{Morandi}, A., {Ettori}, S., \& {Moscardini}, L. 2007, \mnras, 379, 518

\bibitem[{{Motl} {et~al.}(2005){Motl}, {Hallman}, {Burns}, \&
  {Norman}}]{Motl:05}
{Motl}, P.~M., {Hallman}, E.~J., {Burns}, J.~O., \& {Norman}, M.~L. 2005,
  \apjl, 623, L63

\bibitem[{{Muanwong} {et~al.}(2006){Muanwong}, {Kay}, \&
  {Thomas}}]{Muanwong:06}
{Muanwong}, O., {Kay}, S.~T., \& {Thomas}, P.~A. 2006, \apj, 649, 640

\bibitem[{{Muanwong} {et~al.}(2002){Muanwong}, {Thomas}, {Kay}, \&
  {Pearce}}]{Muanwong:02}
{Muanwong}, O., {Thomas}, P.~A., {Kay}, S.~T., \& {Pearce}, F.~R. 2002, \mnras,
  336, 527

\bibitem[{{Muanwong} {et~al.}(2001){Muanwong}, {Thomas}, {Kay}, {Pearce}, \&
  {Couchman}}]{Muanwong:01}
{Muanwong}, O., {Thomas}, P.~A., {Kay}, S.~T., {Pearce}, F.~R., \& {Couchman},
  H.~M.~P. 2001, \apjl, 552, L27

\bibitem[{{Nagai}(2006)}]{Nagai:06}
{Nagai}, D. 2006, \apj, 650, 538

\bibitem[{{Nagamine} {et~al.}(2006){Nagamine}, {Ostriker}, {Fukugita}, \&
  {Cen}}]{Nagamine:06}
{Nagamine}, K., {Ostriker}, J.~P., {Fukugita}, M., \& {Cen}, R. 2006, \apj,
  653, 881

\bibitem[{{Navarro} {et~al.}(1996){Navarro}, {Frenk}, \& {White}}]{Navarro:96}
{Navarro}, J.~F., {Frenk}, C.~S., \& {White}, S.~D.~M. 1996, \apj, 462, 563

\bibitem[{{Navarro} {et~al.}(1997){Navarro}, {Frenk}, \& {White}}]{Navarro:97}
---. 1997, \apj, 490, 493

\bibitem[{{O'Hara} {et~al.}(2006){O'Hara}, {Mohr}, {Bialek}, \&
  {Evrard}}]{OHara:06}
{O'Hara}, T.~B., {Mohr}, J.~J., {Bialek}, J.~J., \& {Evrard}, A.~E. 2006, \apj,
  639, 64

\bibitem[{{Ostriker} {et~al.}(2005){Ostriker}, {Bode}, \&
  {Babul}}]{Ostriker:05}
{Ostriker}, J.~P., {Bode}, P., \& {Babul}, A. 2005, \apj, 634, 964

\bibitem[{{Pearce} \& {Couchman}(1997)}]{Pearce:97}
{Pearce}, F.~R. \& {Couchman}, H.~M.~P. 1997, New Astronomy, 2, 411

\bibitem[{{Pearce} {et~al.}(2000){Pearce}, {Thomas}, {Couchman}, \&
  {Edge}}]{Pearce:00}
{Pearce}, F.~R., {Thomas}, P.~A., {Couchman}, H.~M.~P., \& {Edge}, A.~C. 2000,
  \mnras, 317, 1029

\bibitem[{Perlmutter {et~al.}(1999)Perlmutter, Aldering, Goldhaber, Knop,
  Nugent, Castro, Deustua, Fabbro, Goobar, Groom, Hook, Kim, Kim, Lee, Nunes,
  Pain, Pennypacker, Quimby, Lidman, Ellis, Irwin, McMahon, Ruiz-Lapuente,
  Walton, Schaefer, Boyle, Filippenko, Matheson, Fruchter, Panagia, Newberg, \&
  Couch}]{perlmutter:99}
Perlmutter, S., Aldering, G., Goldhaber, G., Knop, R., Nugent, P., Castro, P.,
  Deustua, S., Fabbro, S., Goobar, A., Groom, D.~E., Hook, I.~M., Kim, A.~G.,
  Kim, M., Lee, J., Nunes, N., Pain, R., Pennypacker, C., Quimby, R., Lidman,
  C., Ellis, R., Irwin, M., McMahon, R., Ruiz-Lapuente, P., Walton, N.,
  Schaefer, B., Boyle, B., Filippenko, A., Matheson, T., Fruchter, A., Panagia,
  N., Newberg, H. J.~M., \& Couch, W. 1999, \apj, 517, 565

\bibitem[{{Pires} {et~al.}(2006){Pires}, {Juin}, {Yvon}, {Moudden}, {Anthoine},
  \& {Pierpaoli}}]{Pires:06}
{Pires}, S., {Juin}, J.~B., {Yvon}, D., {Moudden}, Y., {Anthoine}, S., \&
  {Pierpaoli}, E. 2006, \aap, 455, 741

\bibitem[{{Poole} {et~al.}(2007){Poole}, {Babul}, {McCarthy}, {Fardal},
  {Bildfell}, {Quinn}, \& {Mahdavi}}]{Poole:07}
{Poole}, G.~B., {Babul}, A., {McCarthy}, I.~G., {Fardal}, M.~A., {Bildfell},
  C.~J., {Quinn}, T., \& {Mahdavi}, A. 2007, \mnras, 380, 437

\bibitem[{{Romeo} {et~al.}(2006){Romeo}, {Sommer-Larsen}, {Portinari}, \&
  {Antonuccio-Delogu}}]{Romeo:06}
{Romeo}, A.~D., {Sommer-Larsen}, J., {Portinari}, L., \& {Antonuccio-Delogu},
  V. 2006, \mnras, 373, 1648

\bibitem[{{Ruhl}(2004)}]{Ruhl:04}
{Ruhl}, J., e.~a. 2004, in Presented at the Society of Photo-Optical
  Instrumentation Engineers (SPIE) Conference, Vol. 5498, Millimeter and
  Submillimeter Detectors for Astronomy II., ed. C.~M. {Bradford}, P.~A.~R.
  {Ade}, J.~E. {Aguirre}, J.~J. {Bock}, M.~{Dragovan}, L.~{Duband}, L.~{Earle},
  J.~{Glenn}, H.~{Matsuhara}, B.~J. {Naylor}, H.~T. {Nguyen}, M.~{Yun}, \&
  J.~{Zmuidzinas}, 11--29

\bibitem[{{Sch{\"a}fer} {et~al.}(2006{\natexlab{a}}){Sch{\"a}fer}, {Pfrommer},
  {Bartelmann}, {Springel}, \& {Hernquist}}]{Schafer:06}
{Sch{\"a}fer}, B.~M., {Pfrommer}, C., {Bartelmann}, M., {Springel}, V., \&
  {Hernquist}, L. 2006{\natexlab{a}}, \mnras, 370, 1309

\bibitem[{{Sch{\"a}fer} {et~al.}(2006{\natexlab{b}}){Sch{\"a}fer}, {Pfrommer},
  {Hell}, \& {Bartelmann}}]{Schafer:06b}
{Sch{\"a}fer}, B.~M., {Pfrommer}, C., {Hell}, R.~M., \& {Bartelmann}, M.
  2006{\natexlab{b}}, \mnras, 370, 1713

\bibitem[{{Schmidt} {et~al.}(1998){Schmidt}, {Suntzeff}, {Phillips},
  {Schommer}, {Clocchiatti}, {Kirshner}, {Garnavich}, {Challis}, {Leibundgut},
  {Spyromilio}, {Riess}, {Filippenko}, {Hamuy}, {Smith}, {Hogan}, {Stubbs},
  {Diercks}, {Reiss}, {Gilliland}, {Tonry}, {Maza}, {Dressler}, {Walsh}, \&
  {Ciardullo}}]{schmidt:98}
{Schmidt}, B.~P., {Suntzeff}, N.~B., {Phillips}, M.~M., {Schommer}, R.~A.,
  {Clocchiatti}, A., {Kirshner}, R.~P., {Garnavich}, P., {Challis}, P.,
  {Leibundgut}, B., {Spyromilio}, J., {Riess}, A.~G., {Filippenko}, A.~V.,
  {Hamuy}, M., {Smith}, R.~C., {Hogan}, C., {Stubbs}, C., {Diercks}, A.,
  {Reiss}, D., {Gilliland}, R., {Tonry}, J., {Maza}, J.~e., {Dressler}, A.,
  {Walsh}, J., \& {Ciardullo}, R. 1998, \apj, 507, 46

\bibitem[{{Sehgal} {et~al.}(2007){Sehgal}, {Trac}, {Huffenberger}, \&
  {Bode}}]{Sehgal:07}
{Sehgal}, N., {Trac}, H., {Huffenberger}, K., \& {Bode}, P. 2007, \apj, 664,
  149

\bibitem[{{Shaw} {et~al.}(2006){Shaw}, {Weller}, {Ostriker}, \&
  {Bode}}]{Shaw:06}
{Shaw}, L.~D., {Weller}, J., {Ostriker}, J.~P., \& {Bode}, P. 2006, \apj, 646,
  815

\bibitem[{{Shaw} {et~al.}(2007){Shaw}, {Weller}, {Ostriker}, \&
  {Bode}}]{Shaw:07}
---. 2007, \apj, 659, 1082

\bibitem[{{Sijacki} \& {Springel}(2006)}]{Sijacki:06}
{Sijacki}, D. \& {Springel}, V. 2006, ArXiv Astrophysics e-prints

\bibitem[{{Solanes} {et~al.}(2005){Solanes}, {Manrique}, {Gonz{\'a}lez-Casado},
  \& {Salvador-Sol{\'e}}}]{Solanes:05}
{Solanes}, J.~M., {Manrique}, A., {Gonz{\'a}lez-Casado}, G., \&
  {Salvador-Sol{\'e}}, E. 2005, \apj, 628, 45

\bibitem[{{Spergel} {et~al.}(2007){Spergel}, {Bean}, {Dor{\'e}}, {Nolta},
  {Bennett}, {Dunkley}, {Hinshaw}, {Jarosik}, {Komatsu}, {Page}, {Peiris},
  {Verde}, {Halpern}, {Hill}, {Kogut}, {Limon}, {Meyer}, {Odegard}, {Tucker},
  {Weiland}, {Wollack}, \& {Wright}}]{Spergel:07}
{Spergel}, D.~N., {Bean}, R., {Dor{\'e}}, O., {Nolta}, M.~R., {Bennett}, C.~L.,
  {Dunkley}, J., {Hinshaw}, G., {Jarosik}, N., {Komatsu}, E., {Page}, L.,
  {Peiris}, H.~V., {Verde}, L., {Halpern}, M., {Hill}, R.~S., {Kogut}, A.,
  {Limon}, M., {Meyer}, S.~S., {Odegard}, N., {Tucker}, G.~S., {Weiland},
  J.~L., {Wollack}, E., \& {Wright}, E.~L. 2007, \apjs, 170, 377

\bibitem[{Stadel {et~al.}(1997)Stadel, Katz, Weinberg, \&
  Hernquist}]{Stadel:97}
Stadel, J., Katz, N., Weinberg, D.~H., \& Hernquist, L. 1997, {\tt
  www-hpcc.astro.washington.edu/tools/skid.html}

\bibitem[{{Sunyaev} \& {Zel'dovich}(1972)}]{sunyaev:72}
{Sunyaev}, R.~A. \& {Zel'dovich}, Y.~B. 1972, Comments Astrophys. Space Phys.,
  4, 173

\bibitem[{{Thomas} \& {Couchman}(1992)}]{Thomas:92}
{Thomas}, P.~A. \& {Couchman}, H.~M.~P. 1992, \mnras, 257, 11

\bibitem[{{Thomas} {et~al.}(2002){Thomas}, {Muanwong}, {Kay}, \&
  {Liddle}}]{Thomas:02}
{Thomas}, P.~A., {Muanwong}, O., {Kay}, S.~T., \& {Liddle}, A.~R. 2002, \mnras,
  330, L48

\bibitem[{{Vikhlinin} {et~al.}(2006){Vikhlinin}, {Kravtsov}, {Forman}, {Jones},
  {Markevitch}, {Murray}, \& {Van Speybroeck}}]{Vikhlinin:06}
{Vikhlinin}, A., {Kravtsov}, A., {Forman}, W., {Jones}, C., {Markevitch}, M.,
  {Murray}, S.~S., \& {Van Speybroeck}, L. 2006, \apj, 640, 691

\bibitem[{{Voevodkin} \& {Vikhlinin}(2004)}]{Voevodkin:04}
{Voevodkin}, A. \& {Vikhlinin}, A. 2004, \apj, 601, 610

\bibitem[{{Wechsler} {et~al.}(2002){Wechsler}, {Bullock}, {Primack},
  {Kravtsov}, \& {Dekel}}]{Wechsler:02}
{Wechsler}, R.~H., {Bullock}, J.~S., {Primack}, J.~R., {Kravtsov}, A.~V., \&
  {Dekel}, A. 2002, \apj, 568, 52

\bibitem[{{Wechsler} {et~al.}(2006){Wechsler}, {Zentner}, {Bullock},
  {Kravtsov}, \& {Allgood}}]{Wechsler:06}
{Wechsler}, R.~H., {Zentner}, A.~R., {Bullock}, J.~S., {Kravtsov}, A.~V., \&
  {Allgood}, B. 2006, \apj, 652, 71

\bibitem[{{Weller} {et~al.}(2001){Weller}, {Battye}, \& {Kneissl}}]{weller:01}
{Weller}, J., {Battye}, R., \& {Kneissl}, R. 2001, \prl, 88, 231301

\bibitem[{{White}(2001)}]{white:01}
{White}, M. 2001, \aap, 367, 27

\bibitem[{{White} {et~al.}(2002){White}, {Hernquist}, \& {Springel}}]{White:02}
{White}, M., {Hernquist}, L., \& {Springel}, V. 2002, \apj, 579, 16

\bibitem[{{Zhang} {et~al.}(2007){Zhang}, {Finoguenov}, {B{\"o}hringer},
  {Kneib}, {Smith}, {Czoske}, \& {Soucail}}]{Zhang:07}
{Zhang}, Y.-Y., {Finoguenov}, A., {B{\"o}hringer}, H., {Kneib}, J.-P., {Smith},
  G.~P., {Czoske}, O., \& {Soucail}, G. 2007, \aap, 467, 437

\end{thebibliography}
%\bibliographystyle{../../../papers/astronat/apj/apj}

\end{document}